# Dynamic and rate-dependent yielding in model cohesive suspensions.

Richard Buscall[1,2*], Peter J Scales[2], Anthony D Stickland[2], Hui-En Teo[2] and Daniel R Lester.[3]

[1] MSACT Research and Consulting, Exeter, EX2 8GP, UK.
[2] Particulate Fluid Processing Centre, Dept. Chemical and Biomolecular Engineering, University of Melbourne, Australia 3010.
[3] Dept of Chemical Engineering, Royal Melbourne Institute of Technology, Melbourne, Australia 3001.

* Author to whom correspondence should be sent by email to r.buscall@physics.org

**Abstract**
An experimental system has been found recently, a coagulated $CaCO_3$ suspension system, which shows very variable yield behaviour depending upon how it is tested and, specifically, at what rate it is sheared. At Péclet numbers Pe > 1 it behaves as a simple Herschel Bulkley liquid, whereas at Pe < 1 highly non-monotonic flow curves are seen. In controlled stress testing it shows hysteresis and shear banding and in the usual type of stress scan, used to measure flow curves in controlled stress mode routinely, it can show very erratic and irreproducible behaviour. All of these features will be attributed here to a dependence of the solid phase, or, yield stress, on the prevailing rate of shear at the yield point. Stress growth curves obtained from step strain-rate testing showed that this rate-dependence was a consequence of Péclet number dependent strain softening. At very low Pe, yield was cooperative and the yield strain was order-one, whereas as Pe approached unity, the yield strain reduced to that needed to break interparticle bonds, causing the yield stress to be greatly reduced. It is suspected that rate-dependent yield could well be the rule rather than the exception for cohesive suspensions more generally. If so, then the Herschel-Bulkley equation can usefully be generalized to read $\sigma = \sigma_0 g(\dot{\gamma}) + \sigma_{iso} + k\dot{\gamma}^n$ (in simple shear). The proposition that rate-dependent yield might be general for cohesive suspensions is amenable to critical experimental testing by a range of means and along lines suggested.

## 1. Introduction

An earlier paper [1] described the shear flow of two strongly cohesive suspensions showing highly non-monotonic flow curves, one of which was a 40% v/v suspension of 4.5 μm CaCO$_3$ in water, coagulated by having the pH its 'natural' pH close to the iso-electric point. Here the rheology of the CaCO$_3$ system is examined in more detail with an emphasis on transient behaviour and how it controls the steady state. The effect of solids concentration on the shear rheology will be reported also.

The way in which the original 40%v/v CaCO$_3$ system presented itself as a yield stress liquid was found to depend upon how it was caused to flow [1], as is summarised in table 1 below. With regard to the table, please note that 'CR' denotes 'controlled rate', that 'CS' means 'controlled stress' and that Pe$_0$ is the so-called 'bare' Péclet number $6\pi \bar{a}^3 \mu \dot{\gamma} / k_B T$, where $\bar{a}$ is the mean particle radius, $\mu$ the viscosity of the liquid phase, $T$ is absolute temperature, k$_B$ is Boltzmann's constant and $\dot{\gamma}$ is the shear-rate.

**Table 1: Yield behaviour depends upon test type.**

| | Test protocol | | Behaviour |
|---|---|---|---|
| A | An ascending "staircase" of shear rates in time, all at Pe$_0$ > 1. | CR | Herschel-Bulkley [1]. |
| B | As above but starting from Pe$_0$ << 1 | CR | Non-monotonic flow curve [1]. |
| C | Creep testing at a series of stresses. | CS | Time-dependent yield over a modest range of stress [2]. |
| D | An ascending "staircase" of stresses in time (CS flow curve). | CS | Erratic yield and shear banding [1]. |
| E | As above but with a return down the staircase of stresses. | CS | Hysteresis between ascending and descending branches [1]. |

Table 2 summarises the variation of the apparent yield stress with test type and compares it with a pattern reported earlier by Pham et al. [5] for a weakly-cohesive but very concentrated (60%v/v) non-aqueous dispersion of PMMA particles, depletion-flocculated with dissolved polystyrene. Pham et al. did not see shear-rate dependent yield, but that apart, their variation in apparent yield stress follows a pattern not dissimilar to that seen for CaCO$_3$, qualitatively-speaking.

**Table 2: Approximate variation in apparent yield stress by method compared with that seen by Pham et al. [5]**

(The stress values have been scaled on the largest value measured).

| Method | Pham et al. PMMA [5] $\varphi=0.6$. | CaCO$_3$ [1] $\varphi=0.4$ |
|---|---|---|
| Peak stress on flow start-up @ constant shear-rate. | 1 | 0.5 – 1 (rate-dep$^t$) |
| Strain sweep or staircase. | 0.67 | > 0.5 |
| Stress sweep or staircase. | 0.56 | 0.26 – 0.36 |
| Extrapolation from flow curve. | 0.13 | ~0 |

In [1] it was suggested that modifying or extending the Herschel-Bulkley equation [3] thus could account for the flow curves obtained by controlled rate testing,

$$\sigma = \sigma_s + k_1 \dot{\gamma}^n \Rightarrow \sigma_0 g(\dot{\gamma}) + \sigma_{iso} + k_1 \dot{\gamma}^n \ . \tag{1.1}$$

In eqn 1.1 the yield stress has been split into two parts, a shear-rate dependent part, taken to decrease with increasing shear-rate and to decay to zero at some point, together with a second fixed solid-phase stress term, $\sigma_{iso}$, included to recover Herschel-Bulkley as limiting behaviour at higher shear rates.

The flow curve and the fits are re-plotted in fig. 1 with some additions. Please note also that an error made in the original plot in [1] has been corrected and doing so has changed the position of the falling part of the curve on the abscissa somewhat, it improves the fit to the peak stress on the rising branch also. In the case CaCO$_3$ at 40% v/v, the residual yield stress, $\sigma_{iso}$, could be taken as zero for fitting purposes, although it need not be more generally; indeed, it was substantial for the other suspension described in [1], for example, and it is becomes significant for the CaCO$_3$ system too at concentrations > 40%v/v.

Because of the difficulty of converting angular velocity to shear rate in the case of such a complex flow curve, it was expedient to fit the raw flow curve of stress versus angular velocity in the first instance, using, by analogy to eqn 1.1,

$$\sigma = \sigma_s + k' \Omega^n = \sigma_0 g_{app}(\Omega) + \sigma_{iso} + k' \Omega^n \ . \tag{1.2}$$

This in turn is tantamount to fitting using the apparent, Newtonian shear rate $\dot{\gamma}_N$ thus,

$$\sigma = \sigma_s + k'\dot{\gamma}_N^n = \sigma_0 g_{app}(\dot{\gamma}_N) + \sigma_{iso} + k\dot{\gamma}_N^n \ , \tag{1.3}$$

since $\dot{\gamma}_N$ is proportional to $\Omega$. Such a fit gives a true value for the power-law index $n$ and apparent values for the consistency index and the softening function. The problem of estimating the true shear rate will be addressed here in section 2.

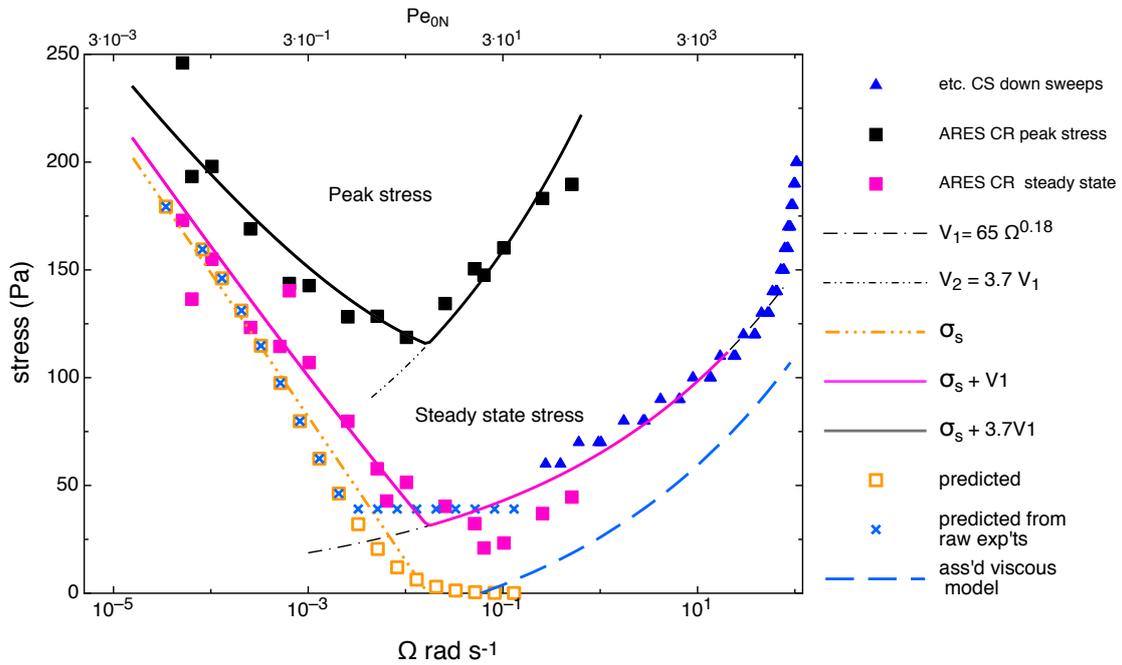

Fig.1 Flow curves re-plotted from [1] for 40 v/v coagulated $CaCO_3$ with a numerical error in [1] corrected, which inter alia improves the peak stress fit. The larger filled squares come from controlled rate measurements. The smaller circles and triangles to the right are from controlled stress testing, stepping the stress downwards from the highest value. The total steady-state stress can be fitted by using the sum of a solid-phase term $\sigma_s = \sigma_0 g(\dot{\gamma})$, assumed to decrease logarithmically, and a power-law viscous term $V_1$: note also that a shear-thickening region at the extreme right [1] has been ignored in the fit. The peak stress measured in start-up can be modelled using the same value of $\sigma_s$ but with a larger viscous term = $3.7V_1$. The crosses and unfilled squares represent predicted values for the solid-phase stress $\sigma_s$ calculated from the strain-softening exponents discussed in section 3.3 using eqn 3.8 (see text for details). The upper axis shows the apparent 'bare' Péclet number calculated from the apparent, or Newtonian shear-rate at the vane, the subscript N denoting this.

It can be seen from fig. 1 that solid phase stress drops to zero logarithmically, thus the overall fit to the stress took the functional form,

$$\sigma_s \approx \sigma_{fit} = k_0 \ln(\dot{\gamma}_N^0 / \dot{\gamma}_N) + k\dot{\gamma}_N^n \tag{1.4}$$

where $\dot{\gamma}_N^0$ is the value of $\dot{\gamma}_N$ at which the fitted stress appears to extrapolate to zero. From the upper axis in fig. 1 it can be seen that $\sigma_s$ does so at an apparent bare $Pe_{0N}$, calculated from the Newtonian shear-rate at the vane (hence $Pe_{0N}$), of order unity. The observation that the solid-phase stress decreases with Pe is unprecedented, so far as we are aware. Koumakis and Petekidis [4], working on the same system as Pham et al. [5], did not see any such effect at values similar $Pe_{0N}$, for example. They did however suggest that an effect of Pe was to be expected, but that it should be seen at very high $Pe_0 \gg 1$. They proposed that it should be controlled by a re-scaled Péclet number, $Pe_{dep} \sim F\, Pe_0$, where $F$ is the magnitude of the dimensionless inter-particle cohesive force. For the $CaCO_3$ suspensions of interest here, $Pe_{dep}$ is conservatively estimated to be at least $10^5\, Pe_0$ (from a consideration of the Van der Waals forces); hence it is clear that Koumakis and Petekidis' expectation is not borne out in practice. Indeed, from the left-hand side of fig.1, it can be seen that softening is already underway at $Pe_{0N}$ values ca. eight orders of magnitude lower. Possible reasons why Koumakis and Petekidis' experimental system did not show softening at $Pe_{0N} < 1$, whereas the $CaCO_3$ system does, will be suggested later.

The scaling rule of Koumakis and Petekidis, $Pe_{dep} \sim F\, Pe_0$, was based on the idea that there is a competition between shear disrupting the local environment, or 'cage' and diffusion and re-bonding trying to re-form it, which is entirely reasonable so far as it goes, of course. It is suspected though, that they might possibly have assumed that the attractive inter-particle force accelerates the rate of re-bonding. It has however long been understood that whereas an attractive force will retard spontaneous escape or debonding in a Brownian system, it will not accelerate capture to any significant extent, the reason being that lubrication forces oppose the attractive force and largely nullify the effect the latter would otherwise have on rate in their absence. Thus, and for example, the rate of fast coagulation of colloidal suspensions is found to be close to ideal Smoluchowski rate and only very weakly dependent upon Hamaker constant, even for order of magnitude variations of the latter, as is illustrated in [23], for example. That the $CaCO_3$ system shows softening on a scale of $Pe_0 \sim 1$, the high volume-fraction not withstanding, means presumably that particle motion on the very short length scales involved in re-bonding is not all that significantly affected by crowding.

Eqn 1.1 suffices to account for much of the behaviour summarised in table 1. What it does not capture is the very variable yielding seen in controlled stress flow curve determination [1]. Liquids with non-monotonic flow curves must shear-band in controlled stress, and in pressure driven flows too, but that of itself does not explain the erratic behaviour seen. By contrast, creep in controlled stress [2] was found to be reasonably reproducible, implying that so would be flow curve determinations in CS mode too, supposing, that is, that the dwell times at each stress were to be made long enough (or, where stress is ramped continuously, where the ramp rate was made slow enough). The implication then is that in CS flow curve determination at strain-rate softening can feed forward to amplify the effect of any small variations in instantaneous strain-rate from one run to the next when the ramp rate is too fast, although quite how that might work has yet to be established.

The paper is organised as follows. In the next section, the problem of estimating the true shear-rate and shear-rate distribution in wide-gap Couette flow will be considered. In the subsequent section stress growth data obtained from flow start up in step-rate will presented. The data will be compared with similar data for other systems taken from the literature, specifically with that of Petekidis et al. [4,5] and Yin and Solomon [6] for systems of much smaller particle size. It will be seen inter alia that outside of the linear region, the $CaCO_3$ system first strain hardens and then strain-softens. In subsequent sections the strain hardening will be analysed in terms of inter-particle forces and the strain softening information will be used to support or justify the fit to the flow curve based on equation 1 and to explore the mechanism of strain-rate softening. This will be shown to be strain-rate dependent strain softening inasmuch that it will be shown that, whereas the $CaCO_3$ system strain softens at all shear-rates, it does so in a way that depends upon the shear-rate.

It will be argued from the data presented here and in [1] that various types of yielding behaviour thought to be disparate hitherto, can be unified or rationalised in terms of Péclet number dependent strain softening of the solid phase stress. This proposition should be very amenable to further testing since it leads to some very distinct predictions regarding the outcome of a number of possible experiments. Some suggestions along those lines will be made and the practical implications of rate-dependent yielding will be discussed briefly too, the key point being that it can cause the one material to behave very differently in pressure-controlled and kinematically-controlled process flows, whereas a simple yield stress liquid would look the same in both.

**2. Estimation of the true shear rate in wide gap Couette flow at controlled rate.**

It is necessary to use wide gaps in the rheometric testing of cohesive suspensions in order to prevent premature yield and slip at the inner surface of the outer cylinder [2]. Here the problem of estimating the true shear rate at the cylindrical surface swept by a vane is considered. The right-hand branch of the flow curve, where the stress increases with rotation rate, presents no problem of course. For example, in the power-law region of the curve, the true shear-rate at the vane is given as usual by,

$$\dot{\gamma}_1 = \dot{\gamma}_{1N} / n \approx 2\Omega / n \, , \qquad (2.1)$$

in the wide gap approximation, where *n* is the power-law index, as in eqn 1. Note that the subscript '1' here denotes "at the vane" (whereas elsewhere in the paper any mention of 'shear-rate' will always refer to that at the vane implicitly, by default). For other than power-law behaviour, *n* can be replaced by it local value $d\ln\sigma/d\ln\Omega$, of course, provided that the latter is positive, and hence there is no difficulty in determining the true shear-rate for 40% CaCO$_3$ in, say, the shear-thickening regime seen at the highest angular velocities too.

Matters become more problematic on the left-hand branch where the stress decreases with increasing rate. It is not too difficult to see what is going on here though, as the liquid is behaving as if it has a yield stress, but one that changes from point to point. Thus, at any one point, the liquid can be thought of as behaving as a Herschel-Bulkley liquid with its own particular yield stress. Were that set of yield stresses to be known then it would be a straightforward matter to calculate the true shear-rates That implies though, that at each rotation rate $\Omega$, there must be unique and different value of the ratio of the true shear rate to the apparent Newtonian shear-rate; call that set of ratios $f(\Omega)$. The calculation of shear-rate at the vane and in the gap for a Herschel-Bulkley liquid is straightforward when the yield stress is a constant [7]. When it is not, it is simply not possible to proceed without some means of decomposing the total stress into solid-phase and viscous parts, a priori, no matter how provisional.

In [1] and in corrected re-plot, fig. 1 here, the flow curve for 40% v/v CaCO$_3$, has been fitted by assuming that the viscous power-law implied by the right-hand branch of positive slope underpins the left-hand branch of negative slope too, as shown. The fit (cf. eqn 1.2) thus specifies a split of the total stress into solid-phase and viscous components at each point and hence it can be used to calculate the shear-rate correction function $f(\Omega)$ consistent with that fit. One would prefer to calculate $f(\Omega)$ in an absolute, model-free way, but that is simply not possible when the slope $d\ln\sigma/d\ln\Omega$ is negative.

Since $\sigma_s \approx \sigma_0 g_{app}(\Omega)$ is a monotonic function of $\Omega$ on the falling left-hand branch, then it must be a unique function of the apparent and true vane shear-rates, thus,

$$\sigma_s(\dot\gamma_1) = \sigma_s(f(\Omega)\dot\gamma_{1N}), \qquad (2.2)$$

We do not know $\sigma_s(\dot\gamma_1)$ of course, we only know its dependence upon $\Omega$ and thus $\dot\gamma_{1N}$, as given by the fit in eqn 1.4. Hence the task is to calculate the correction function $f(\Omega)$.

The rotation rate is related to the true shear-rate distribution in the cylinder as standard by,

$$\Omega = \int_{x_0}^{1} d\ln x \ \dot{\gamma}(\sigma[x]) = \int_{x_0}^{1} d\ln x \ \dot{\gamma}(\sigma_1 / x^2); \quad x = r/R_v, \tag{2.3}$$

where the dimensionless cut-off radius $x_0$ is taken to be, either, the dimensionless yield radius, or, the outer cylinder radius, whichever the smaller. From eqn 1.4, the total fitted stress at the vane, $\sigma_1$, can be written as a function of $\sigma_s$ thus,

$$\sigma_1 = \sigma_s + k \left[ \dot{\gamma}_{1N}^0 \exp\left(-\frac{\sigma_s}{k_0}\right) \right]^n \tag{2.4}$$

Since the shear rate at the vane is proportional to the rotation rate, everything else being equal, it must also be the case that,

$$\Omega = \Omega_N f(\Omega) \tag{2.5}$$

where $\Omega_N$ is a notional rotation rate calculated from eqn 2.3 by using the known $\dot{\gamma}_N$ in place of the unknown $\dot{\gamma}$. From eqns 2.3 and 2.4 then,

$$\Omega_N = \int_{x_0}^{1} d\ln x \ \dot{\gamma}_N(\sigma_1 / x^2) = \int_{x_0}^{1} dx \left[ \frac{k \left[ \dot{\gamma}_N^0 \exp(-\sigma_s / k_0) \right]^n + \sigma_s(1 - x^2)}{k x^{2+n}} \right]^{1/n} \tag{2.6}$$

The integration of the RHS of eqn 2.6 can be performed analytically to produce,

$$\Omega_N = \frac{n}{2} \left[ \begin{array}{c} \left(\dfrac{\sigma_s}{k}\right)^{1/n} {}_2F_1(\tfrac{-1}{n}, \tfrac{-1}{n}, \tfrac{n-1}{n}, 1) \\ -\left(\dfrac{k\left[\dot{\gamma}_N^0 \exp(-\sigma_s / k_0)\right]^n + \sigma_s}{k}\right)^{1/n} {}_2F_1(\tfrac{-1}{n}, \tfrac{-1}{n}, \tfrac{n-1}{n}, x_0^{-2}) \end{array} \right], \tag{2.7}$$

where the function $_2F_1$ is the ordinary (or Gauss') hypergeometric function defined by the series,

$$_2F_1(q, b, c, z) = \frac{(b)_j}{(c)_j} \frac{z^j}{j!} + \sum_{1}^{\infty} \frac{(q)_j (b)_j}{(c)_j} \frac{z^j}{j!};$$

$$\text{with,} \quad (q)_j = q(q+1)\ldots(q+j-1). \tag{2.8}$$

The cut-off distance appearing in the second term of eqn 2.9 and elsewhere, the scaled yield radius, is given by,

$$x_0 = \sqrt{\frac{k}{\sigma_s}\left[\dot{\gamma}_N^0 \exp(-\sigma_s/k_0)\right]^n + 1}, \qquad (2.9)$$

It is evident then that the shear-rate correction function $f(\Omega)$ can be calculated by comparing $\Omega_N$ with the true angular velocity $\Omega$, since, from eqn 2.5, $f(\Omega) = \Omega/\Omega_N$. The latter is plotted as the continuous line in fig. 2. It can be seen that the difference between the corrected and notional Newtonian shear-rate increases with decreasing rotation rate, as would be expected, and that it becomes very large indeed at low rates. A good part of the difference comes from the viscous power law, given that $1/n = 5.56$, as is shown by the dashed line. The remainder comes from the decrease in yield radius $x_0$ with decreasing rotation rate.

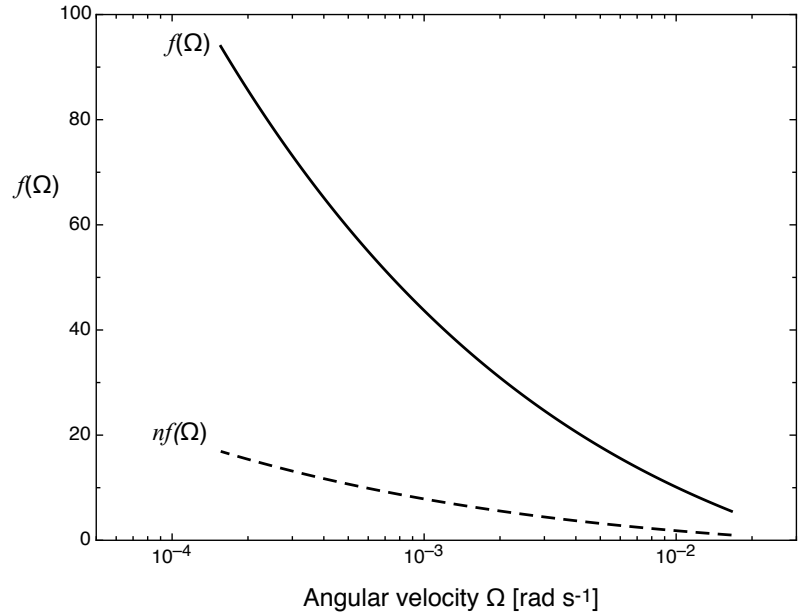

Fig. 2 The shear rate correction function from eqns 2.5 to 2.7. The upper curve shows the total correction, the dashed line that with the power-law contribution of $1/n$ factored out.

The shear-rate profiles in the cylindrical gap are plotted in fig. 3. They are normalised on the value at the vane (at $x = 1$). The flow becomes ever more confined to a region close to the vane as the solid-phase stress increases, and the rotation rate and hence the yield radius decreases, again as is to be expected. It can be seen that the flow is largely confined to a region with a thickness of less that half the vane radius even in the absence of a yield stress because of the low value of $n$.

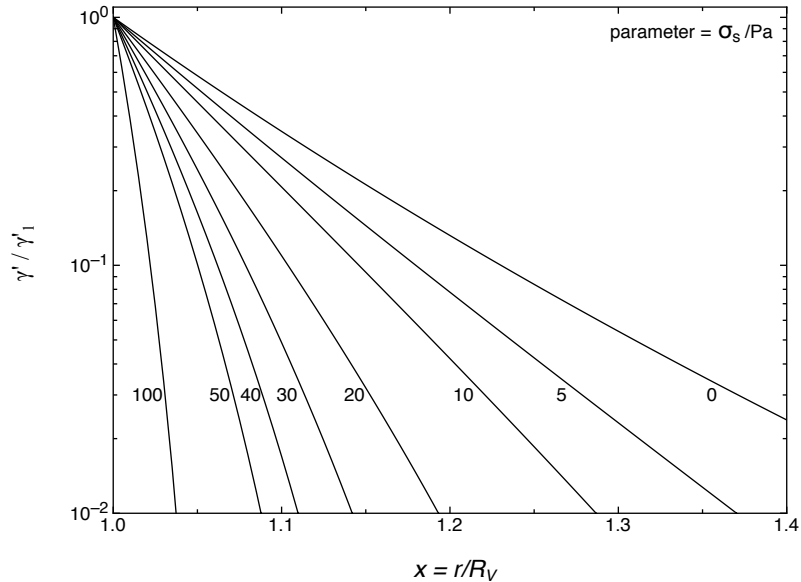

Fig. 3 normalised shear rate distributions away from the vane for various values of the solid-phase stress, and with the curve labelled zero showing the limiting power-law behaviour. In all other cases the shear-rate falls away to zero at the yield-radius, even if the downturn is not all that apparent in this plot over 2 decades of rate.

The reader might reasonably think that the corrected shear-rates should not be taken too seriously without some sort of validation, given that they are dependent on the fitting model used to force a decomposition of the total stress into solid phase and viscous parts. Here though, the fact that the peak and the steady state-stresses can both be fitted using the same solid-phase stress term, and by changing the consistency index $k$ only, gives confidence. Could one do better more generally and, say, validate the results by means of further measurements? A direct way of doing so, given that $\sigma_s = \sigma_1 / x_0^2$, would be to attempt measure the yield radius at each rate by means of visualization, although doing so with sufficient accuracy might be difficult in practice, not least because the yield radii in the region of interest are small, as can be seen from fig. 3.

The solid-phase stress is plotted against the Newtonian and corrected shear-rates in fig. 4. The dotted line merely serves to show again that the total correction comprises a change in shape and a power-law shift. The overall effect on the flow curve is thus two-fold: the whole curve is shifted towards higher rates by a factor $1/n$ and the left-hand, solid phase stress dominated branch is rendered much steeper. That the corrected shear rate appears to be double-valued in the stress at low stress is almost certainly an artefact coming from the curve fits, since the corrected curve is very sensitive to the precise nature of the fit here. The scatter in the experimental data made iteration to eliminate this small artefact scarcely worthwhile.

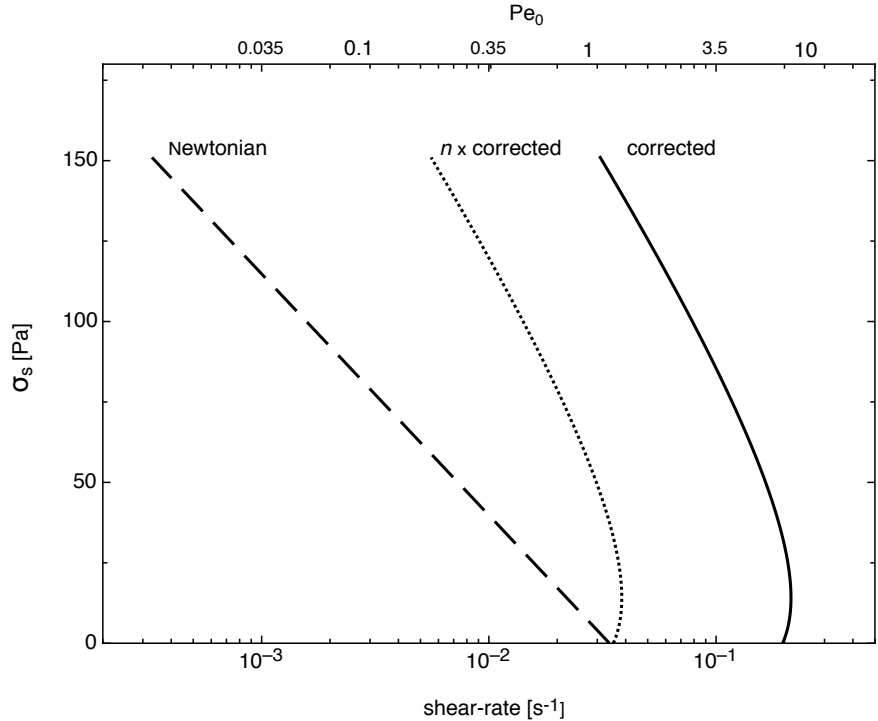

**Fig. 4** solid phase stress ($\sigma_s$ in fig.1) plotted against apparent and corrected shear-rate. The dotted line shows the corrected shear-rate multiplied by the power-law index. That the corrected shear rate is double valued in the stress is almost certainly an artefact coming from the curve fits as the corrected curve is very sensitive to the precise details of the fit here. The scatter of the experimental data however, makes iteration, to eliminate this, scarcely worthwhile.

The upper scale, which shows some values for the bare $Pe_0$ calculated from the corrected shear-rate, shows that softening takes place at $Pe_0 \sim 1$. It is important to emphasize though that the curve shown is simply the corrected relationship between steady-state shear-rate and yield stress. It is not necessarily the notional inverse, the causal dependence of yield stress on deformation rate, even though it must be some sort of reflection of it. The reason for saying this is simply that a steady state cannot be established unless or until the solid-phase stress sets, implying that the shear rate must still be transient when the former becomes fixed. Indeed, the fact that the peak stress can be fitted using the same yield stress function $\sigma_s \approx \sigma_0 g_{app}(\Omega)$ confirms this point. The matter will be discussed at length later in the context of transient behaviour, but in the mean time, suffice to say that the actual rate-dependence of the yield stress must be bracketed by the 'Newtonian' and corrected curves, not that that is saying all that much, perhaps.

It needs to be emphasized too that the calculations above apply to controlled rate rheometry only. In the controlled stress, the region of negative slope is inaccessible since such a suspension must shear-band there. The fits to the flow curve alone do not then allow the shear rate distribution to be calculated, in general, because the location of the boundary between the stagnant and flowing bands is unknown in the absence of a quantitative shear-banding criterion. The position is not of necessity always quite so hopeless though when the viscous stress contribution happens to be strongly shear thinning, as it is in the case of the $CaCO_3$ suspensions. When that is so, it is easy to show that the shear-rate at the vane depends only very weakly upon the location of the shear band boundary, or it does, that is, provided that the latter is not very close to the vane, as it cannot be on the right-hand branch of fig. 1 for example. More generally though, the rule fixing the location of the band boundary needs to be known. This would especially so when the viscous power-law happens to be nearer to unity, as it was the case for the other shear-banding material described in [1].

**3. Stress growth in step shear-rate.**

The first sub-section below will present specimen stress-growth curves, will compare them with similar data from the literature and then analyze the strain softening seen at intermediate strains and its rate-dependence. The second sub-section will analyze the strain hardening seen in the early stages of stress growth prior to the softening in terms of inter-particle forces. The third sub-section will contend that rate-dependent softening controls the yield stress and strain.

*3.1 Stress growth curves: their time scaling and comparison with data from the literature.*

All the measurements on $CaCO_3$ suspensions were made using cruciform vanes in wide cylindrical cups as in [1]. A summary of essential details is given in an appendix. Figs. 5a and 5b show stress growth curves for 40%v/v $CaCO_3$ for various values of the apparent Newtonian shear-rate. It was shown in [1] that the curves could be largely scaled in terms of time by plotting against the engineering strain $\gamma = \dot{\gamma} t$, to a give a family of stress-strain curves. The engineering strain covers a very wide range from $<10^{-4}$ to ca. 5, and it exceeds unity well before a steady state is approached. In recognition of that, $\gamma$ will be replaced here by an appropriate or equivalent scalar derived from the Hencky (or natural) strain tensor, this being given by [8],

$$\gamma_H = 2\ln\left(\gamma/2 + \sqrt{1+\gamma^2/4}\right) = \ln\left(1+\gamma^2/2 + \gamma\sqrt{1+\gamma^2/4}\right) \qquad (3.1)$$

This reduces to $\gamma$ at small strain, as it should, and it differs little from $\gamma$ until the latter becomes of order unity.

Fig. 5a shows stress strain curves for lower shear-rates where the peak stress decreases with increasing shear-rate, as indicated by the arrow, whereas fig. 5b shows data for higher shear-rates where the peak rises again. The peak position at $\gamma_H \sim$ 0.5+/- 0.2 varied a little, albeit not systematically, but only by a little, hence the time scaling by $\dot{\gamma}_{1N}$ is very good, given that the latter ranges over 4 decades or so. By contrast, had the corrected shear-rate plotted in fig. 2 been used then instead, then the peak position in strain would have been found to vary by a factor of twenty or so. This is seen as the first piece of evidence that the shear-rate at the vane when yield occurs must be much closer to $\dot{\gamma}_N$ than it is to the steady-state value.

Rather similar behaviour was seen at 25, 30 and 43% v/v, except that on the average the strain at the peak increased with decreasing concentration: the variation at each concentration was such that it was not possible to pin the dependence down precisely, although it looked to be approximately reciprocal, with the typical strains at peak for 30 and 25% being more like 1 and 1.5, respectively.

Overall, the stress growth curves for 40%v/v could be divided into five regions starting with a linear region at very low strain. Or, rather, it could, given that there must have been a linear region below a strain of ca. 3e-5, since the data were very noisy in that region. Above ca. 3e-5 strain-hardening was seen over something less than a decade strain, followed in turn by a very extended region of strain-softening between ca. 2e-4 and 0.5 (typically). After the peak (region 4), the stress decayed to a noisy or fluctuating steady state; the impression gained being that the system hunted around some mean or typical value of the stress in the steady state indefinitely.

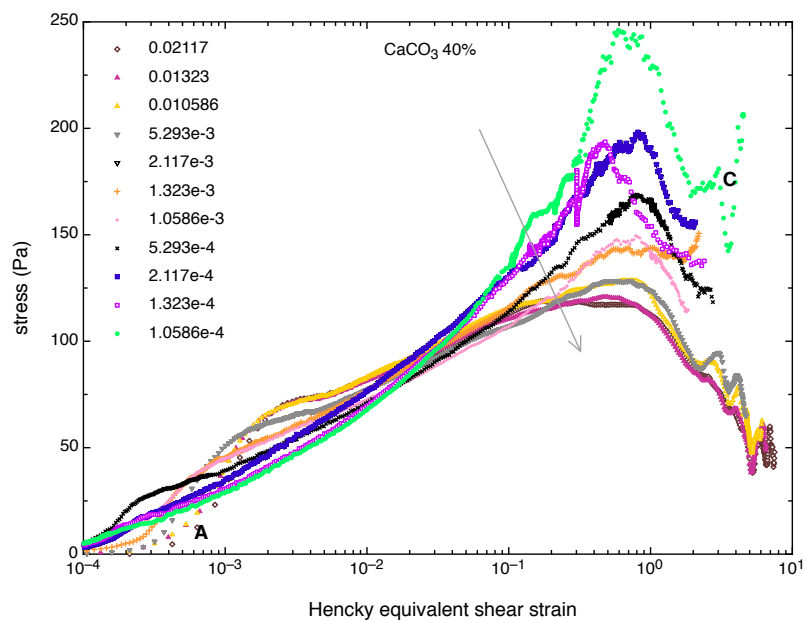

**Fig. 5a.** Stress growth at step strain-rate for 40%v/v CaCO$_3$. Data re-plotted from [1]. The abscissa is the Hencky equivalent shear strain magnitude given by $\ln\left(1 + \gamma/2 + \gamma\sqrt{1 + \gamma^2/4}\right)$ where $\gamma = \dot{\gamma}t$ is the engineering shear strain.

The parameter is the apparent or Newtonian shear rate at the vane. The plot shows the data for the lower shear rates where the peak <u>decreases</u> with shear-rate (whereas 5b, below, shows the curves for the higher shear-rates where the peak grows again). The spread at 'A' is a result of inertia. 'C' draws attention to the erratic approach to steady state: it was found that the steady state was somewhat noisy at all rates as if the system hunts around it.

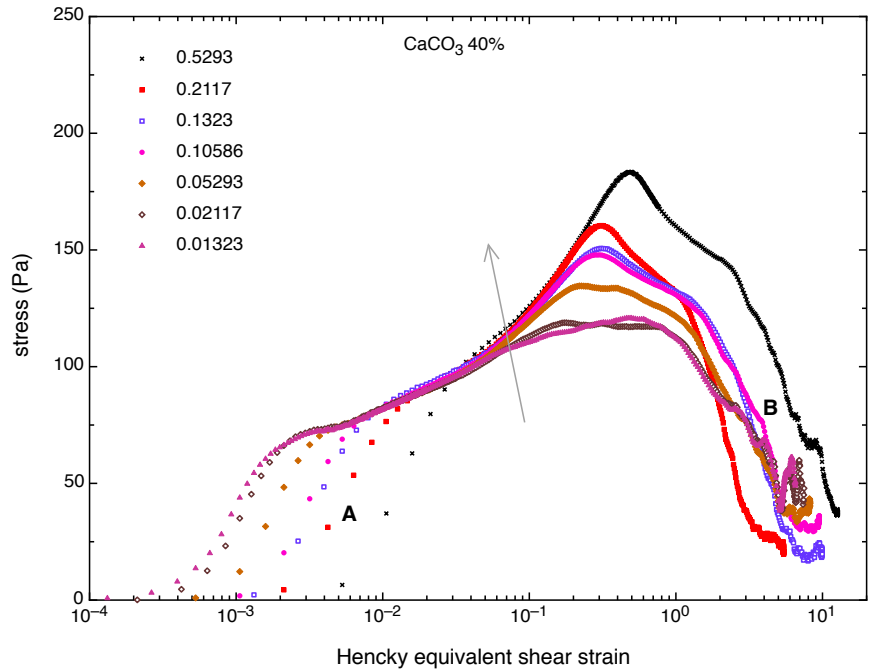

**Fig. 5b.** Stress growth at step strain-rate for 40%v/v CaCO$_3$ at higher shear-rates where the peak grows again. 'B' - The time scaling seems to change as the viscous stress becomes dominant (data replotted from [1]).

Petekidis et al. [4, 5], in their work on depletion flocculated PMMA particles, refer to two 'yield strains, a lower yield strain where the solid phase starts to soften, and a second strain where the stress peaks. It is however perhaps preferable to speak of a 'softening strain' and a 'yield strain', given that their system remained solid until the second strain. Softening at a lower strain followed by yield at a strain ~ 1 has been seen more widely in creep testing too [e.g. 2, 8-11]. The characteristic softening strain is found to correlate with the expected range of the inter-particle force [e.g. 2,4-11], implying that cohesive particulate gels strain more or less affinely until bonds are obliged to break. The current data support this idea too.

The present stress growth curves differ qualitatively from those of Petekidis et al. [4,5] in one very important respect, inasmuch that their peak stress only increased monotonically with shear-rate, even though the ranges of Pe$_{0N}$ covered are comparable. Their stress growth curve at low Pe$_{0N}$ for 60%v/v PMMA from [4] is re-plotted in the lower half of fig. 6, where the stress has been scaled on the linear shear modulus. It is compared with the corresponding curves for 30 and 40%v/v CaCO$_3$ at similar Pe$_{0N}$. Also shown are data for 20% v/v flocculated silica of yet smaller particle size (86 nm diameter) from Yin and Solomon [6]. Their experiment was somewhat different in that they carried out stress relaxation experiments at a series of strains, the points plotted here being the stress at a very short time delay $t$-$t_0$ = 0.01s, where $t_0$ is the rise time used to attain the set strain at a fixed imposed rate of strain. The data should however be identical to that which would have been obtained from step-strain rate stress growth tests at the same shear-rate, provided that no relaxation has occurred over the time $t$-$t_0$. Also shown is the same experiment performed for CaCO$_3$ in order to confirm that this does indeed gives the same results as the step strain-rate stress growth experiment, as it should. The upper half of fig. 6 shows the scaled stress divided by strain to give an integral strain softening function $h(\gamma_H) = \sigma / \gamma_H$ ('integral' to distinguish from the alternative way of defining such a function as $h'(\gamma_H) = d\sigma / d\gamma_H$).

It can be seen that the stress for silica peaks at much lower strain than that for PMMA or CaCO$_3$. In each case, though, the softening strain is close to the characteristic 'bond' strain indicated by the arrows at the top of fig. 6, the latter being the theoretically estimated ratio of the surface-to-surface separation of maximum inter-particle force to the particle diameter.

The silica shows power-law strain softening with an exponent of -1.4. The CaCO$_3$ exhibits power-law behaviour likewise, but with a smaller exponent of ca. -0.7. In the case of the PMMA, the 'bond strain' is much closer to the yield strain, making it more difficult to extract an exponent, although the slope looks to be very similar to that for the CaCO$_3$. A possible reason why the softening function deviates from power-law near the peak will be discussed below. It is clear from the comparison of the PMMA and CaCO$_3$ data with the silica data, that softening exponent determines the yield strain. Should the softening exponent be > -1 (i.e. if its magnitude is < 1), then the yield strain is ~1, whereas if the exponents is < -1 then the stress peaks at the softening, or, (characteristic) bond strain. Taken together, the data shown in fig. 6 provide a hint that the yield strain could be volume-fraction dependent, perhaps, given that the silica has the lowest. As we shall show though, it turns out to be both volume-fraction and Péclet number dependent in the case of CaCO$_3$, since the yield strain will be seen to revert to the bond strain at higher Pe too, regardless of the concentration.

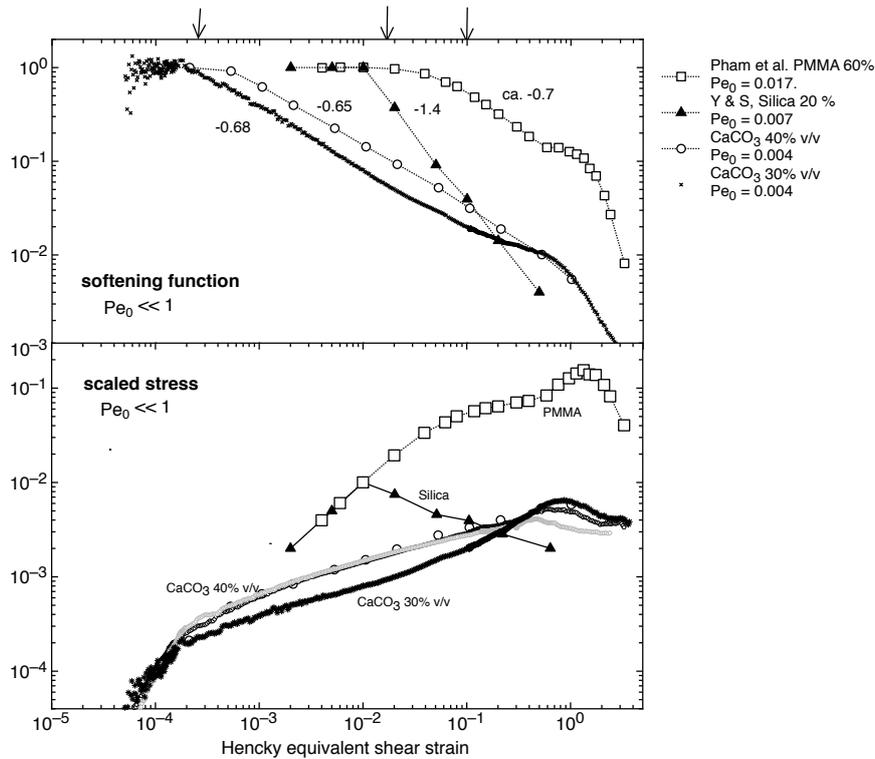

**Fig. 6.** Dimensionless shear stress, scaled on the shear modulus, (lower plot) and integral softening function (upper) versus Hencky strain for 40 and 30% v/v CaCO$_3$ compared with data for 60% v/v depletion-flocculated PMMA from Pham et al. [5] and 20% v/v incipiently flocculated silica from Yin and Solomon [6]. The Pe$_{0N}$ values are all low and comparable. The arrows above the upper horizontal axis indicate the characteristic 'bond' strains estimated from the scaled inter-particle force (n.b. were everything else to be equal this would simply be inversely proportional to particle size, whereas the long-range depletion flocculation reverses the order for the PMMA relative to the silica). The data for silica were obtained from a series of stress-relaxation tests, as were the open circles for 40% v/v CaCO$_3$. The remaining data come from stress growth curves in step-strain rate. The two tests agree for 40% v/v CaCO$_3$, as they should.

Some sample strain softening function plots for CaCO$_3$ are shown in figs. 7a and b. The curves were sensibly power-law over an extended range of strain in most cases, although shear-rate had an effect since inertia delays stress growth at higher shear-rates and limits the power-law range. The exponents are plotted against Pe$_{0N}$ in fig. 8. They tend to decrease from a low Pe value of ca. 0.68+/-0.03 to a plateau value that decreases with concentration, overall, even if those for 30% and 40% appear to coincide more or less. The arrows at the top show the Pe$_{0N}$ numbers pertaining to the data of Pham et al. [5] and Yin and Solomon [6] plotted in fig. 6.

At 25%v/v the exponent drops somewhat below -1 above a Pe$_{0N}$ ~ 0.4, causing the yield strain to revert from a value ~ 1 to that of the bond strain, or thereabouts. Thus, the yield strain was Pe dependent, at this volume-fraction at least, such that yield ceased to be cooperative at higher Pe. An attempt was made to reduce the concentration further in order to see whether the yield strain would then revert to the bond-strain at all Pe, as the data of Yin and Solomon hint that it might, but, unfortunately, CaCO$_3$ suspensions then sediment too rapidly to be tested with any confidence. One would need to use smaller or less dense particles in order to go to 20% v/v or below, presumably.

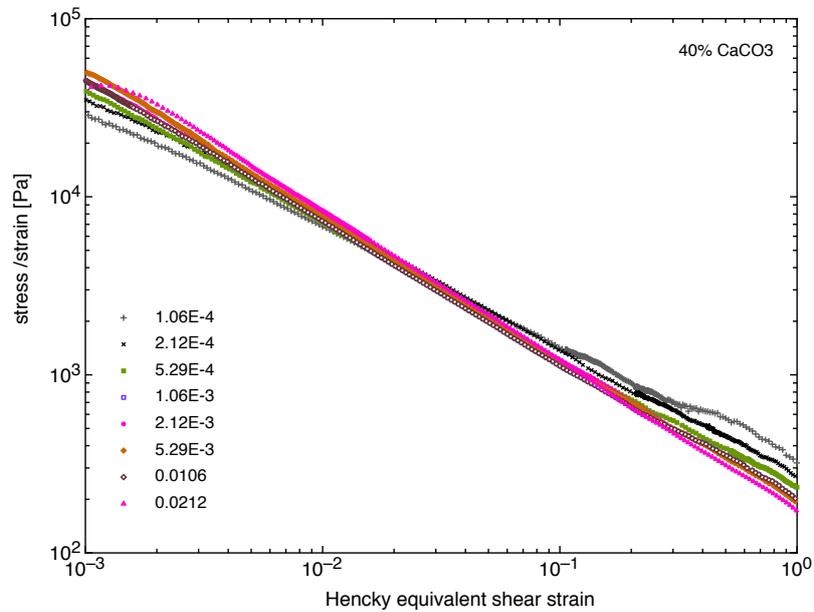

**Fig. 7a.** Some representative strain softening curves for 40% v/v CaCO$_3$ at various shear-rates. Curves at other shear rates have been omitted for clarity.

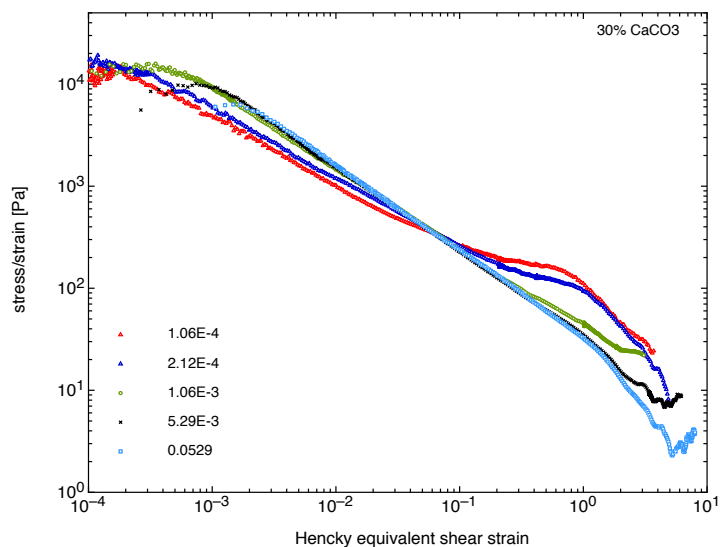

**Fig. 7b** Some representative strain softening curves for 30% v/v CaCO$_3$ at various shear-rates. Curves at other shear rates have been omitted for clarity. The tendency for the curves to show a peak at the left at higher shear rates is a result of inertia. It happens at all concentrations but becomes intrusive at lower shear-rates for lower concentrations, for obvious reasons.

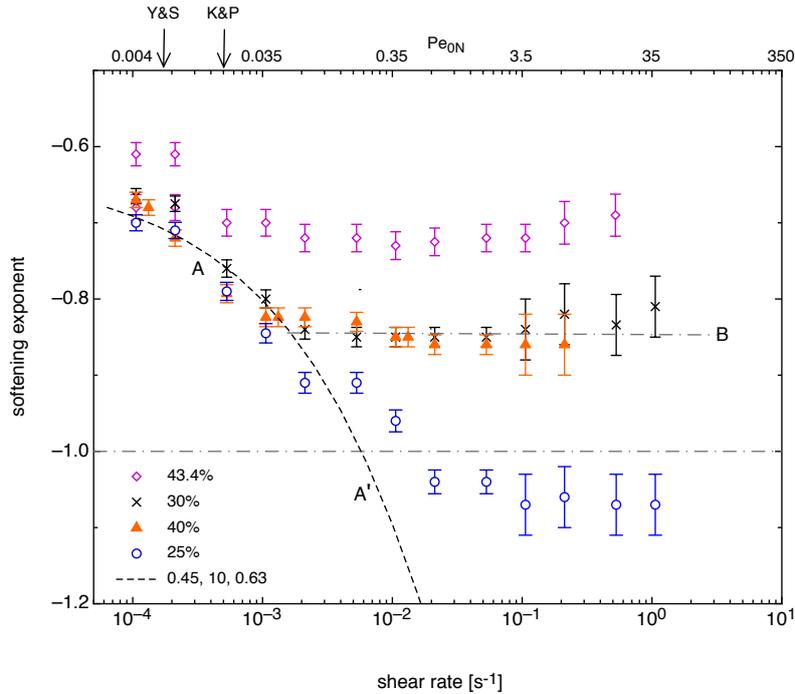

**Fig. 8**. Strain softening exponent versus Newtonian shear rate and $Pe_{0N}$ for $CaCO_3$ at four concentrations. The fits A-A' and A-B will be used later to model the flow curves. The arrows at the top show the $Pe_{0N}$ numbers pertaining to data of Pham et al [5] and Yin and Solomon [6] shown in fig. 6. It was however evident from the behaviour at yet higher shear-rates, that the data go through broad maxima. They do so because inertia pushes the power-law region close to the strain ~ 1 stress peak, where the viscous stress is significant. Between them, inertia and viscosity distort the data at higher shear-rate. It is thought that the true exponent controlling the rate-softening of the solid-phase stress probably follows a curve more like A-A' than, say, A-B: that there is an apparent concentration dependence at higher shear-rates is thought to be a consequence of the viscous stress contribution to the total stress, the apparent exponents having been extracted from plots of total stress.

Koumakis and Petekidis [4], following Pham et al. [5], varied volume-fraction for the depletion-flocculated PMMA system. It is not possible to extract a strain-softening exponent from their data, as the curves did not show a well-defined power-law region. Because of this, the stress-growth curves themselves are re-plotted in fig. 9. It can be seen that the stress peak at higher strain disappears at a volume-fraction of ca. 0.25. Note that these are low $Pe_{0N}$ data and thus comparable with those for $CaCO_3$ etc.

Overall then, the data presented and reviewed here suggest that yield is cooperative with yield strain ~1 at low Pe and for volume-fractions > 0.25, whereas simply breaking bonds suffices below 0.25 and at Pe ~ 1. Further experimental data will be need to confirm the generality of this picture or otherwise, of course. Given that the systems used by Yin and Solomon (6) and Petekidis et al. (4,5) were very much smaller in particle size and strength of attraction than is the $CaCO_3$ system (cf. table 3), it would be most instructive to work with other systems of intermediate particle size perhaps; with particle radii in the range of 250 to 1000 nm diameter, say.

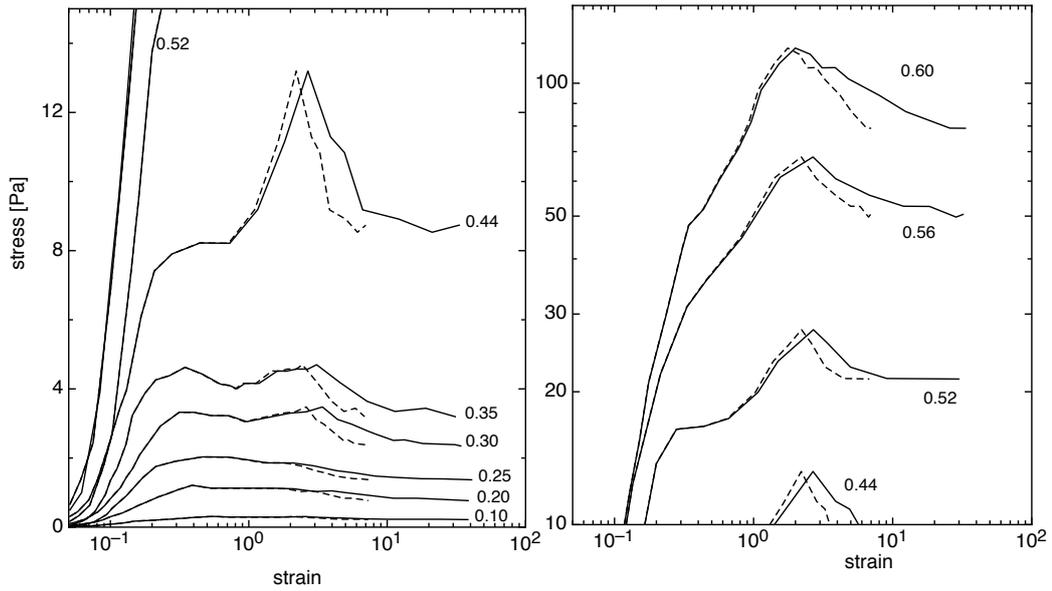

**Fig. 9.** Approximate re-plots of the stress growth data of Koumakis and Petekidis [4] for various volume-fractions, the LH graph showing the behaviour at lower volume-fraction, the log-log plot on the RHS the higher. The continuous lines are against nominal strain; the broken lines against Hencky equivalent shear strain. Dropping the concentration below 0.35 causes the peak at strain ~ 1 first to turn into a noisy plateau and then to move to lower strain.

**Table 3: A comparison of the $CaCO_3$ suspension with those of Petekidis et al. [4,5] and Yin and Solomon [6].**

| Sample | Particle radius $R$/nm | $\varphi$ | Well-depth $-U/k_BT$ | Scaled range $\delta/2R$ | Reference |
|---|---|---|---|---|---|
| Non-aq. silica | 40 | 0.20 | ~20 | ~0.012 | [6] |
| Non. Aq. PMMA | 130 | 0.60 to 0.15 | ~20 | ~0.08 | [4,5] |
| Aq. $CaCO_3$ | 2250 | 0.43 to 0.25 | ~500 | <0.0002 | [1,2], this. |

It should be borne in mind that the exponents plotted in fig. 8 were obtained from plots of the total stress, whereas, were it possible, one would plot the solid phase stress only, since yield involves softening or melting of the solid phase. That the viscous stress contribution to the total stress is significant at larger Pe, even at the yield strain, is evident from the shear-rate dependence of the peak stress. That being the case, then it could be that Petekidis et al. [4,5] only ever saw an increase in peak height with rate simply because the viscous stress happened to be much larger at all Pe in their system, relatively speaking. Their liquid phase, comprising PS in cis-decalin, would certainly have been more viscous, by an order of magnitude perhaps (they do not quote a figure, but the PS used to cause depletion flocculation was at is overlap concentration, or thereabouts). It could also be the case that the solid-phase and viscous stresses scale differently with particle size (etc.) for that system too.

It is suspected that the viscous stress contribution causes the exponents obtained from plots of total stress plotted in fig. 8 to appear to saturate at a plateau. The true dependence of the exponent governing the softening of the solid phase stress on Pe is thought to look more like curve A-A′ than, say, curve A- B. This idea will be tested in section 3.3, but, first, the early stages of stress growth up to the characteristic bond or softening strain will be analysed in terms of interparticle forces.

### *3.2 Early stage growth and strain hardening.*

An unusual feature of the stress-strain curves for $CaCO_3$ is that they show strain hardening at very small strains $< 10^{-4}$ [1]. The aim here is to attempt to account for that by calculating the shape of the early stress-strain curve theoretically from the inter-particle force law. This will be done using a modest extension of the approach described by Zwanzig and Mountain [13] and elaborated more recently by Chateau and Pasol [14]. Their calculations were concerned with small strains and so will we be, except that the problems is more subtle for particles as large as the $CaCO_3$, than it is, say, for molecules and nanoparticles since seemingly minuscule strains can now elicit an anharmonic response from the inter-particle bonds, simply because the strain is concentrated in the bonds, given that the ratio of the unperturbed inter-particle separation to particle diameter is so small. In order to account for this microscopic non-linearity the expansion of the force law will be taken way beyond its first derivative where it is normally terminated [13,14]. The calculations shown here will differ from earlier work [13,14] in that respect only, although some detail will be provided for those not familiar with that and related works.

The key assumption needed in the absence detailed information on the microstructure will be that the pairwise radial distribution function is a delta-function at the interparticle separation of zero-force. Because the $CaCO_3$ particles were coagulated, they can be thought of as being trapped in a deep potential well resulting from the sum of the Van der Waals potential and some unknown but very short-range repulsion. It is possible to calculate the Van der Waals potential and force, given that the Hamaker constant is quoted as being ca. 3.5 $k_BT$ [12] and this in turn implies that the well depth must be several hundred $k_BT$, making the delta-function approximation a reasonable one, arguably. The repulsion will be modelled here by means of a $10^{th}$ power potential in the distance of closest approach $H=r-D$, where $r$ is the centre-to-centre separation between two particles and $D$ the particle diameter. The $10^{th}$ power is an arbitrary choice of course, except that the idea is to place the well at a plausible surface-to-surface separation of 1nm, with the position of maximum force of close by. The repulsive strength was then simply chosen to make the equilibrium separation $H_0$ = 1 nm. The use of alternative but sensible choices does not change any of the predictions, qualitatively, whereas the consequences in quantitative terms will be discussed in due course.

Equation 5 from Pasol and Chateau [14] gives the stress derived from the pairwise inter-particle force $\mathbf{F(r)}$ between nearest neighbours for a statistically homogenous suspension as,

$$<\sigma> = -\frac{n^2}{2}\int_V dV(\mathbf{r})g(\mathbf{r})\mathbf{r}\otimes\mathbf{F(r)} \qquad (3.2)$$

where the lower-case inter-particle separation vector $\mathbf{r}$ refers to the deformed configuration, whereas, upper case $\mathbf{R}$ will refer to the undeformed reference configuration. The circled cross is the direct vector product, as usual, i.e. the matrix product of $\mathbf{r}$ and the transpose of $\mathbf{F}$.

Pasol and Chateau then obtained the following result, their eqn 6,

$$<\sigma> = -\frac{n}{2}\int_{V_0} dV \left(\mathbf{A}:\varepsilon\otimes\mathbf{F}+\mathbf{R}\otimes\frac{d\mathbf{F}}{d\mathbf{R}}\cdot\mathbf{A}:\varepsilon\right)g_0 + \mathbf{R}\otimes\mathbf{F}\left(\frac{dg_0}{d\mathbf{R}}\cdot\mathbf{A}:\varepsilon\right), \qquad (3.3)$$

$$\mathbf{A} = \frac{d\mathbf{R}}{d\varepsilon},$$

by expanding the force in powers of $\mathbf{R}$ and truncating it to first order, which we do not want to do, of course. The macro-strain may be small in the early stages, but in our case the force becomes non-linear in $R$, at strains of order $H_0/D$: We are thus interested tiny macro-strains but finite micro-strains. The other approximations they make, of isotropy and central forces, so will we: isotropy is assumed because the effective volume fraction is high and central forces because the focus is on the initial non-linearity, where central forces can be expected to dominate. The force law, taken to be the sum of Van der Waals plus the short-range power-law repulsion described above, can be written as,

$$F(H) = \frac{DA_H}{24H_0^2}\left[\left(\frac{H_0}{H}\right)^2 - \left(\frac{H_0}{H}\right)^{m+1}\right] \tag{3.4}$$

where, $A_H$ is the Hamaker constant for $CaCO_3/H_2O/CaCO_3$ (~3.5 $k_BT$), and where the first term is VdW attraction and the second is the power-law repulsion. It can be seen that using the equilibrium separation $H_0$ as a parameter obviates the need to specify the strength of the power law repulsion explicitly. In order to proceed it will be assumed that $m=10$ and $H_0 = 1$nm as mentioned above.

It will further be assumed that $g(\mathbf{r}) = g_0(R)$ and that we are dealing with simple shear, with the shear plane being the x-z plane. The above assumptions then allow the following substitutions to be made,

$$\frac{d}{d\mathbf{R}} \to \frac{1}{R}\frac{d}{dR}\mathbf{R} \; ;$$

$$g(\mathbf{r}) \to g_0(R) \to \delta(R+H_0-D) \; ; \tag{3.5}$$

$$\varepsilon \to \dot{\gamma}t\,\mathbf{i} \; ; \mathbf{A}:\varepsilon \to \mathbf{R}\cdot\varepsilon \to \dot{\gamma}t\,\mathbf{i}\cdot\mathbf{R}$$

also, $\mathbf{F} \to \frac{1}{R}\frac{dU}{dR}\mathbf{R}$, where $U$ is the inter-particle potential. Introducing the simplifications and specialisations above into eqn 3.1 then leads to the following expression for the shear stress,

$$<\sigma_{zx}(\dot{\gamma}t)> \;= k_e \int_0^{2\pi}\int_0^{\pi} d\phi\,d\theta\; F(R+\delta R)\sin(\theta)\cos(\theta)\sin(\phi) \tag{3.6}$$

where $\delta R = \dot{\gamma}t\,R\sin(\theta)\cos(\theta)\sin(\phi)$ and $k_e = \frac{3z_m\phi}{4\pi R^2\phi_m}$. In the latter identity the mean coordination number at $\phi$ has been taken implicitly to be $z_m\phi/\phi_m$, where $z_m$ is that of the characteristic local packing, taken to be RCP.

The result above in eqn 3.6 amounts to an average of the projection of the inter-particle force in the shear direction over all orientations, in effect.

For a simple force law like eqn 3.3 above, the integrals can be performed analytically in principle, although the result runs to many pages. A more manageable approach then is to expand the force as a Taylor series, but to high enough order in *q* to capture the asymmetry of the potential. The integrations then just produce a series of numerical coefficients independent of the nature of force law. It was found necessary to go to double figures in the order *q* as the series converges only quite slowly, although only terms odd in the strain are needed, the even terms being null, hence

$$<\sigma> \approx k_e \left[ (\dot{\gamma}tR)C_1 \frac{d^2U}{dr^2}\bigg|_R + (\dot{\gamma}tR)^3 C_3 \frac{d^4U}{dr^4}\bigg|_R \ldots + (\dot{\gamma}tR)^q C_q \frac{d^{q+1}U}{dr^{q+1}}\bigg|_R \ldots \right] \quad (3.7)$$

For force law in eqn 3.4, the derivatives evaluated at $H_0$ are given by,

$$\frac{d^q F}{dr^q}\bigg|_{R, q=\text{odd}} = k_2 H_0^{-q} \left( -(q+1)! + \sum_{j=1}^{q}(q+j-1) \right); \quad k_2 = \frac{DA_H}{24H_0^2} \quad (3.8)$$

Hence the stress can be re-written as a sum of terms comprising the strain in the inter-particle bond raised to the power *q* thus,

$$<\sigma> \approx k_e k_2 \sum_{q \text{ odd}} \chi_q (\dot{\gamma}tR/H_0)^q \quad (3.9)$$

The coefficients $C_q$ and $\chi_q$ in eqns 3.6 and 3.8 are given in table 4 for *q* up to 13.

**Table 4: values of the first seven coefficients in eqns 3.6 and 3.8.**

| q | | Coefficients of $(\dot{\gamma}tR)^q d_q F(r)_R \left( = (\dot{\gamma}tR)^q d_{q+1}U(r)_R \right)$ | | Coefficients of $(\dot{\gamma}tR/H_0)^q$ |
|---|---|---|---|---|
| | q!C_q = | /q!= | C_q | χ_q |
| 1 | $\pi^2/8$ | 1.233701 | 1.233701 | 11.10331 |
| 3 | $(3\pi)^2/8^3$ | 1.73489e-1 | 2.89148e-2 | 48.92384 |
| 5 | $(5\pi)^2/2(16)^3$ | 3.01196e-2 | 2.50997e-4 | 90.26856 |
| 7 | $(35\pi)^2/(128)^3$ | 5.76509e-3 | 1.14387e-5 | 1120.73637 |
| 9 | $(63\pi)^2/2(256)^3$ | 1.16743e-3 | 3.21712e-8 | 1078.33036 |
| 11 | $(231\pi)^2/2(1024)^3$ | 2.45241e-4 | 6.14381e-11 | 864.97565 |
| 13 | $(429\pi)^2/2^2(2048)^3$ | 5.28645e-5 | 8.4895e-14 | 604.79500 |

Fig. 10 below compares the experimental early stage growth with predictions made to 5th (q=9) and 7th (q=13) order in the force law (6th and 8th order in the potential) and where the predicted stress has been scaled to match the correct order of magnitude. An alternative, perhaps, would have been to adjust $H_0$ and the repulsive power-law index to obtain a match, the position on the strain axis, being determined the former only, to a very good approximation, whereas the magnitude depends overtly upon both. It can be seen, however, that the starting guess of $H_0 = 1$nm gets the strain scale about right inasmuch that it predicts strain hardening above a strain of ca. 3e-5. Furthermore, an equilibrium separation of ~1nm might well be thought plausible, given that the interface must be carbonate ion rich and hydrated.

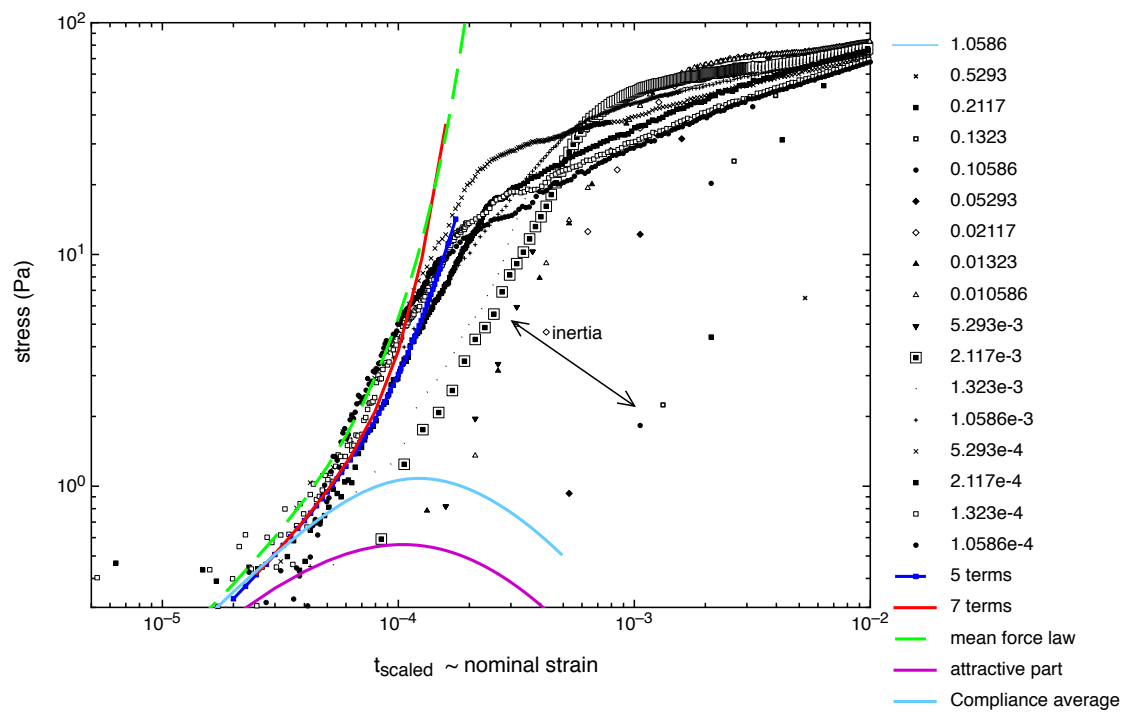

**Fig.10**. The early stage stress growth for 40 %v/v $CaCO_3$ showing the strain hardening between linear and strain softening regions. Inertia disrupts the strain-rate scaling of time to strain at higher strain-rates. For guidance please note that the stress is increasing at a rate of ca. strain raised to the 2.7th power in the strain-hardening region, roughly speaking. The curves are compared with predictions made from the interparticle force law given by eqn 3.4 using eqns 3.5 to 3.9. Also shown are the scaled means of the inter-particle force. Two means are shown, these being parallel (affine or strain) and series (compliance) averages of the stretching and compressional forces. The stretching or attractive part alone is shown too. Its position accounts for the transition from strain hardening to strain softening.

It seems then that the asymmetry of inter-particle potential, together with the concentration of the strain in the interparticle bonds by the factor $D/H_0$, suffice to account for the strain-hardening naturally. Changing the somewhat ad hoc values of $H_0$ and $m$ used does not affect this conclusion: making the repulsion steeper merely accentuates the strain hardening somewhat whereas making it very much softer is simply inconsistent with strong cohesion. The key point here is that in order for the attraction to be very strong, the equilibrium separation has to be on a molecular scale and hence tiny on the scale of the particle size. By contrast, the VdW force is much longer range, making it inevitable that the net interparticle potential must necessarily be highly asymmetric about $H_0$, and that is all that is needed in the model to give strong strain-hardening near the 'bond-strain'; it is thus seen as an inevitable consequence of $D/H_0 \ggg 1$.

The calculations above assume affine displacement and hence they give a strain-average over the microstructure, an upper bound to the stress-strain law. All that has really been done, in effect, is to average the interparticle force over all orientations, hence an approximate calculation can be made by just adding the attractive (stretched) and repulsive (compressed) contributions to the force, given that numbers of bonds in tension and compression can be considered to be equal on the average. The broken line in fig. 10 illustrates this idea. The attractive, or stretch, part is shown too. The latter peaks at ca. 1e-4 where strain softening takes over. By the expedient of simply adding the stretch and compression forces, it is possible to perform an approximate calculation of the lower bound or compliance average of the stress too, as shown also. This looks rather like an amplified version of the attractive branch, not too surprisingly. The early stage strain hardening seen in the experimental data, then, suggests that deformation is affine at small strains. That is not to say deformation continues to be affine at larger strains, even though the excellent time-scaling of the peak stress could perhaps be taken to indicate that it might do so approximately up to strain ~1 at least.

*3.3 strain-softening exponents underpin yield.*

That the peak stress increased with shear-rate at the higher rates suggests that the viscous part of the total stress had grown substantially by $\gamma$ ~1, even though the flow is still far from steady state there. In the case of the steady-state, it was possible to separate the solid-phase stress and viscous stress by means of a plausible fit, as it was too for the peak stress and hence at $\gamma$ ~1. One would like to do the same for all $\gamma$ <1 in order to better define the rate-dependence of the solid phase stress in that region. Possible means of doing so experimentally will be mentioned in section 4 later, whereas in this section an attempt to do so inductively will be attempted using the working hypothesis that the solid phase stress becomes constant at the yield strain. The fits to the peak and steady state stresses in fig. 1 suggest this of course.

That hypothesis leads immediately to the following simple prescription for the solid-phase stress:

$$\sigma_s \approx \sigma_{bond}\left(\frac{\gamma_{cage}}{\gamma_{bond}}\right)^{[1+m(Pe_{0N})]} \quad ; \quad (\gamma = \gamma_H) \quad (3.10)$$

where $m(Pe)$ is the rate-dependent softening exponent, $\sigma_{bond}$ is the value of the stress at the characteristic bond strain, i.e. that at the onset of strain softening, and where the strain is the equivalent Hencky strain defined in eqn 3.1, as has been indicated to the right of eqn 3.10. It is important to note that the exponent in eqn 3.10 is not necessarily the same as that derived from the total stress, as plotted in fig. 8, rather, it is the underlying, true softening exponent for the solid phase stress and the aim here is to try gain a better handle on it. Equation 3.10 then, allows a prediction to be made of the solid phase stress in fig. 1, given a prescription for the Pe dependence of the softening exponent.

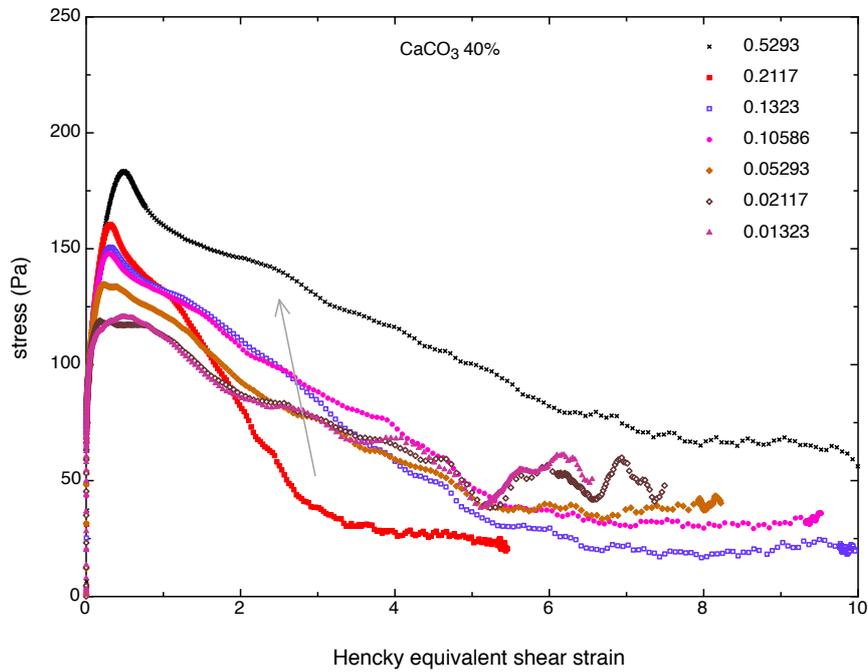

**Fig. 11** Stress growth for 40% v/v CaCO$_3$ on a linear scale to show approach to steady state.

Both lines A-A' and A-B in fig. 8 will be used for this purpose in order to probe the idea that the apparent plateau to the right is an artefact; a consequence an ever increasing viscous component. The Line A-A' was generated using the following expression,

$$m(Pe_{0N}) = -0.63 - 1.37 \left( \frac{\alpha Pe_{0N}}{1 + \alpha Pe_{0N}} \right)^{\beta}$$

(3.11)

$$\alpha = 0.165 \quad \beta = 0.45$$

Such a detailed form might seem a little fanciful or gratuitous perhaps, except that it was sought to give the exponent low and high Pe limits. The resulting predictions are however insensitive to the precise value of the upper limit, provided that it is significantly <-1: hence, an alternative would have been to make the high Pe limit just a little lower than the exponent seen by Yin and Solomon [6] of -1.4, perhaps, rather than the value of -2 chosen, although for the present purposes doing so would make no discernable difference. Line A-B was encoded simply by using eqn 3.11 up to the straight line B shown in fig.8, and then the latter thereafter.

The resulting predictions for the solid phase stress are plotted in fig. 1, from which it can be seen that that made using line A-A′ is close enough to the solid-phase stress implied by curve fitting to make the hypothesis that the solid phase stress is fully-developed at the yield strain very plausible, and to suggest too perhaps that line A-A′ is indeed closer to the true rate dependence of the softening exponent for m ~ -1 than is A-B. On the other hand, the prediction made from line A-B might be thought to look reasonable too, perhaps, given the scatter in the experimental stress values. There is however a problem with this latter prediction though, inasmuch that it cannot be used to fit the whole flow curve without adopting an unphysical form for its accompanying viscous stress; one that vanishes to zero at finite shear-rate, as shown. Hence line A-A′, or something like it, is much to be favoured over A-B.

The reader may feel that we have pushed the interpretation of limited data quite hard here and certainly more data is needed in order to prove the picture developed. The point to bear in mind though is that model, although provisional, is consistent with every aspect of the experimental data, so far as it goes; there are no contradictions or anomalies insofar as we can tell.

This section will be concluded by looking at the concentration dependence of the peak stress. Fig. 12 depicts that of the peak stress at the lowest shear-rate used. It can be fitted by a power-law with an exponent a little above 4, as shown, implying that the peak stress tracks the linear elastic modulus in respect of concentration dependence, given that an exponent of 3.5 to 4.5 has been seen very widely for the modulus of particulate gels. Fig. 13 compares the effect of Newtonian shear-rate and thus $Pe_{0N}$ on the peak stress for each concentration. The rather large scatter not withstanding, these data are consistent with the idea that true strain softening exponent is Pe-dependent but concentration independent, so far as they go, inasmuch that they look as if they might be following a common curve on the left, where they are falling. The rise on the right-hand side becomes larger as the concentration increases, as might be expected, supposing that it is viscous. In saying that fig. 13 is consistent with the idea that the softening exponents are concentration-independent above, it was meant at volume-fraction above 0.25, since the data of Yin and Solomon [6] and Koumakis and Petekidis [4] tend to suggest that cooperative yield at strain ~1 looks to be lost completely at a volume-fraction just below that, or they do supposing that they conform to the one larger picture. The data for $CaCO_3$ at 0.25 suggested also that the cage was fragile there.

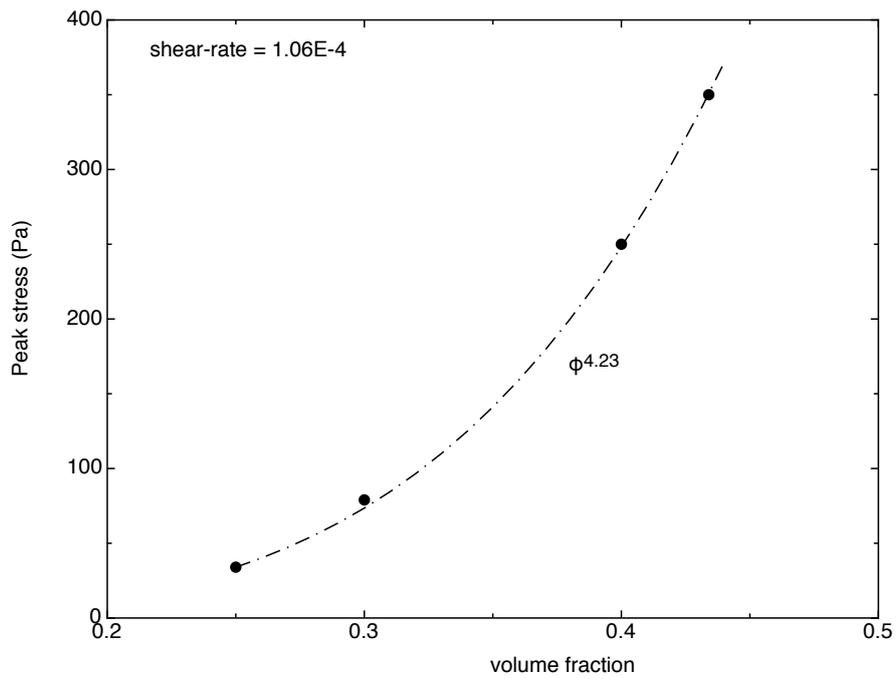

**Fig. 12** Peak stress at low Pe versus volume fraction. The line is a power-law fit.

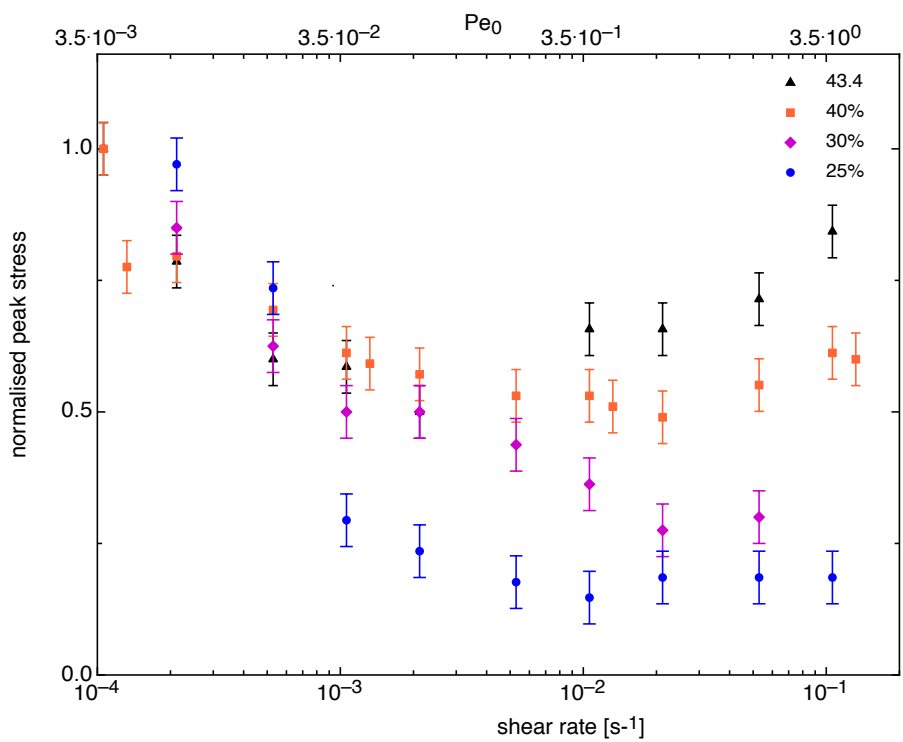

**Fig. 13** Peak stress versus apparent (Newtonian) shear rate for four concentrations. The descending branch looks as if it might have a common dependence before the viscous stress overwhelms the solid-phase part at higher shear rates.

3. **Some observations on stress relaxation, LAOS, creep and creep recovery.**

Other types of dynamic test could perhaps help separate the elastic and viscous stress at strains <1. Furthermore, the model suggested here makes some specific predictions in respect of what might be seen in certain other tests.

The first prediction concerns creep testing. Yield in creep is time-dependent [2, 9-12]. Furthermore, there is often not a sharp yield stress, but a yield range of stress, the existence of which is a manifestation of retarded viscoelasticity below the yield point [2]. Since the gel behaves as a viscoelastic solid prior to yield, the instantaneous strain-rate in creep decreases monotonically with time initially, and in the yield stress range of stress, it drops close to zero before then turning up again as yielding takes place [2, 9-12]. In the yield range then, the strain-rate just prior to yield is small, which, from the current model, would suggest that the yield strain in creep should then be the cage strain, of order one, and that is just what is seen in practice [e.g. 2, 9-12].

Let us turn now to stress relaxation after step strain. An idealised stress relaxation would involve stepping the strain from zero to the set value instantaneously at time $t=0$, but that is impossible in practice of course. In practice the strain is normally increased at some constant set rate to give a rise time $t_0 = \gamma_{set} / \dot{\gamma}_N$. In the absence of strain-rate dependent strain-softening, the relaxation curve obtained at each set strain should be independent of $t_0$ and $\gamma_{set}$, inertial effects not withstanding, whereas strain-rate dependent strain-softening should cause the results to be strongly dependent upon the rate and thus $t_0$. This is exactly what is seen for $CaCO_3$, as is shown in fig.15. At strains intermediate between the bond and yield strains, the suspension relaxes like a viscoelastic solid at low rates and more like a liquid at high rates.

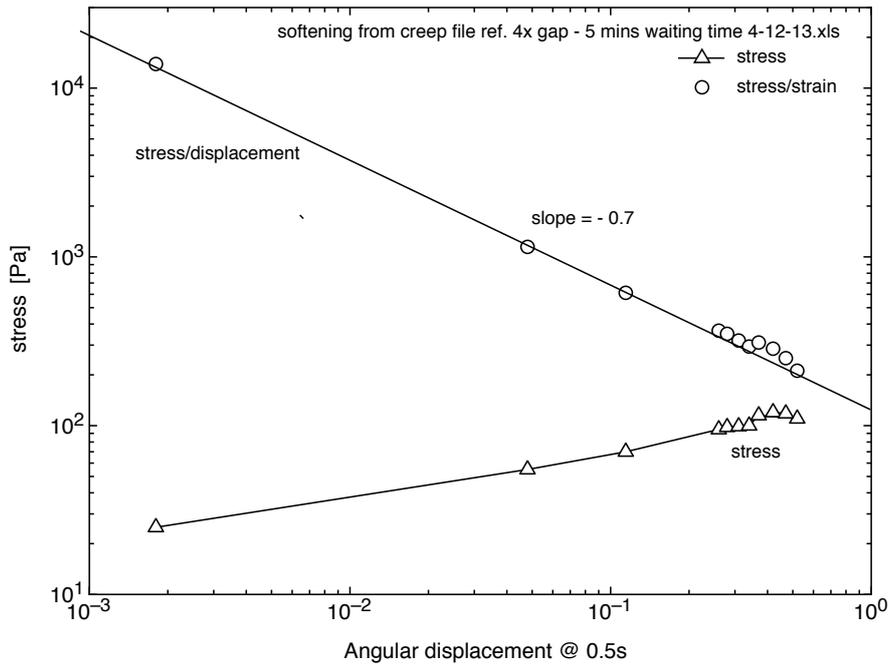

Fig. 14 strain softening at negligible strain rate from creep [2,9] for $CaCO_3$. The exponent of -0.7 is similar to the low Pe values in fig. 8 obtained from step-strain rate, as would be expected.

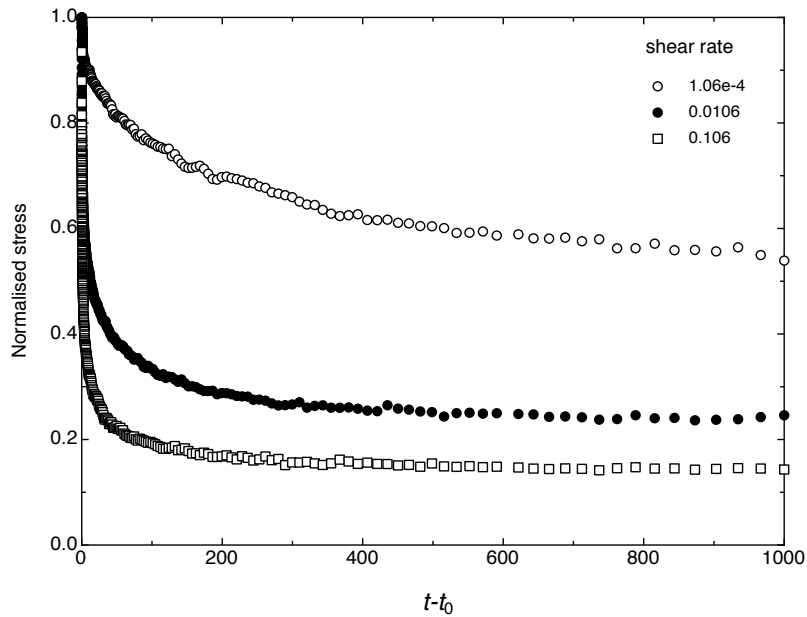

Fig. 15 Stress relaxation for 40% v/v $CaCO_3$ at a strain of 0.106 for three different values of the strain-rate used to achieve that strain.

There is much information to be had from stress relaxation tests in principle. In practice though it can be difficult to obtain good data for very small strains and at long times, since various external sources of mechanical noise can cause spurious relaxation. The data of Yin and Solomon [6] are remarkably clean, even though it is much easier to work at their relatively large strains, than it is near, say, the bond strain of the $CaCO_3$ system. They set a benchmark nevertheless, hence similar work on CaCO$_3$ at low strains and long times in progress, obtaining the data shown in fig.15 at a strain ~ 0.1 being less demanding.

Another way to resolve the split between elastic and viscous stresses in the softening region would to perform creep recovery tests and compare the resulting recovery curves with the preceding creep curves [9], something that is less often done than just creep itself, creep recovery being very difficult to do well, since artifacts such as baseline drift, that can be buried in the creep curves, become obvious in recovery. LAOS too could have a particular use in the present context: viscoelastic materials, be they soft solid or liquids, become more solid-like as the frequency is raised, of course, in the ordinary run of things. The same need not be true necessarily though when there is rate-dependent strain softening, since then the gel should become more liquid-like with increasing frequency over strain-softening region of strain. Hence LAOS frequency sweeps could well offer another useful way of fingerprinting rate-dependent softening and of separating elastic and viscous stresses in the softening region of strain-rate.

## 5. Discussion.

### *5.1 General discussion and suggestions.*

A suspension has been found that shows rich yield behaviour, inasmuch that it yields differently depending upon how it is tested, or, to put that another way, it demonstrates in one system a range of behaviours that in totality have only seen in disparate systems hitherto. For example, it shows all of the features seen by Pham et al. [5] and Koumakis and Petekidis [4], together with more exotic behaviour like non-monotonic flow curves and shear banding. All of this behaviour appears to be a manifestation of having a strain-rate dependent yield stress and the latter can in turn be accounted for in terms of strain-rate dependent strain softening.

A more graphic summary might be to say that whereas yield is cooperative at low strain-rates and Pe << 1, for volume-fractions > 0.25 at least, strain-rate induced melting causes bond-breaking alone to suffice at higher rates; those corresponding to Pe$_0$ ~ 1 or more. The implication is that whereas affine deformation will break bonds at small strains, perikinetic (or Brownian) re-bonding can then rebuild the local structure, so long as Pe$_0$ << 1, causing yield to be postponed to strain ~ 1, whereas, at higher Pe, the rate of bond breaking is just too fast. Taken together, the current data and that of Pham et al. [5] and Koumakis and Petekidis [4], suggest that much the same picture could well apply from the jamming concentration down to volume-fractions as low as 0.25, perhaps, and for a very wide range of particle size, and interaction strength and range. Further work on intermediate particle sizes and in the lower volume-fraction range is needed in order to test this conjecture and challenge the picture.

Our model system has a number of virtues. The large particle size makes a wide and interesting range of Pe accessible; it also separates the bond and cage strains by several orders of magnitude. That the liquid phase is water and not, say, a viscous polymer solution, has the effect that the viscous stress does not dominate everything at low Pe, allowing one to see what the solid-phase stress, or rate-dependent yield stress, is doing. The disadvantages of our $CaCO_3$ model system are, first, that it is very opaque, preventing optical and scattering probes of structure, and, secondly, that it is not possible to get much below a volume-fraction of 0.25 because of settling.

Pham et al. [5] and Koumakis and Petekidis [4] too report a stress peak at ~ 1 which they attribute to cage melting at a strain ~1, except that their peak stress increased with increasing shear-rate implying that it was actually viscous in origin (which is not to say that the fact that the viscous stress too peaks at strain ~ 1 is not somehow associated with cage melting). It should however be possible to construct a modified version of their system but one with lower liquid phase viscosity, either, say, by using lower molecular weight PS to effect depletion flocculation, or, even better perhaps, using reverse micelles, e.g. AOT micelles. It should be possible to increase the cohesion strength by such means too, the well depth by a factor of three, perhaps, and force by a decade or more.

The data presented to date, which include a small amount from creep and stress relaxation, are consistent with one picture based on rate-dependent strain softening. The latter concept explains inter alia why the yield strain in creep is widely or normally seen to be ~1. For the future, stress relaxation, creep recovery and, perhaps, LAOS, used advisedly, are of potential interest as ways of disentangling loss and storage in the softening region, something that needs to be done in order to determine the rate dependence of the softening exponents more precisely. All of these tests difficult to perform well and meaningfully, on systems as strain sensitive as $CaCO_3$ especially, suggesting that they could or should be complementary.

This part of the discussion will be concluded by asking why Koumakis and Petekidis [4] did not see non-monotonic flow curves etc. for the PMMA system (cf. their fig. 12). It could just be that their system was just very different and not prone to rate-dependent softening, for whatever reason. That does not have to be the case though, since an alternative possibility is that the balance of solid-phase and viscous stresses was different and such that viscous stress was dominant enough to mask any underlying rate-dependence of yielding. After all, their dispersions were very different in several respects (cf. table 3); in particle-size, the range and strength of the inter-particle force and the liquid phase viscosity. For their liquid phase, Koumakis and Petekidis used a liquid phase comprising a solution of polystyrene, of MW ~ $10^5$, at its overlap concentration in cis-decalin. That makes it of interest to see what effect an order of magnitude increase in viscosity would to be expected have on the flow curve for 40% v/v $CaCO_3$ system. It might be possible perhaps to do so experimentally by adding corn syrup, but in the meantime and in the absence of experimental data, the obvious thing to do is to take the fitted flow curve in fig. 1 and simply scale the consistency index $k$ and then shift the curve on the abscissa so as to match Péclet numbers.

The effect of doing so by factors of up to fifteen is shown in fig. 16, by way of illustration. The exercise suggests that any increase in viscosity > 10 would disguise the strain-rate softening, everything else being equal. It is just possible then that Koumakis and Petekidis could well have seen more complex behaviour had they used low molecular weight polystyrene or some other means of flocculation. Their yield stresses were relatively small, except that these too would have increased had the molecular weight of the PS been lower, since the well depth would have been larger too. Some obvious experiments, likewise involving ICI Paints' beautiful PMMA particles [22], are suggested.

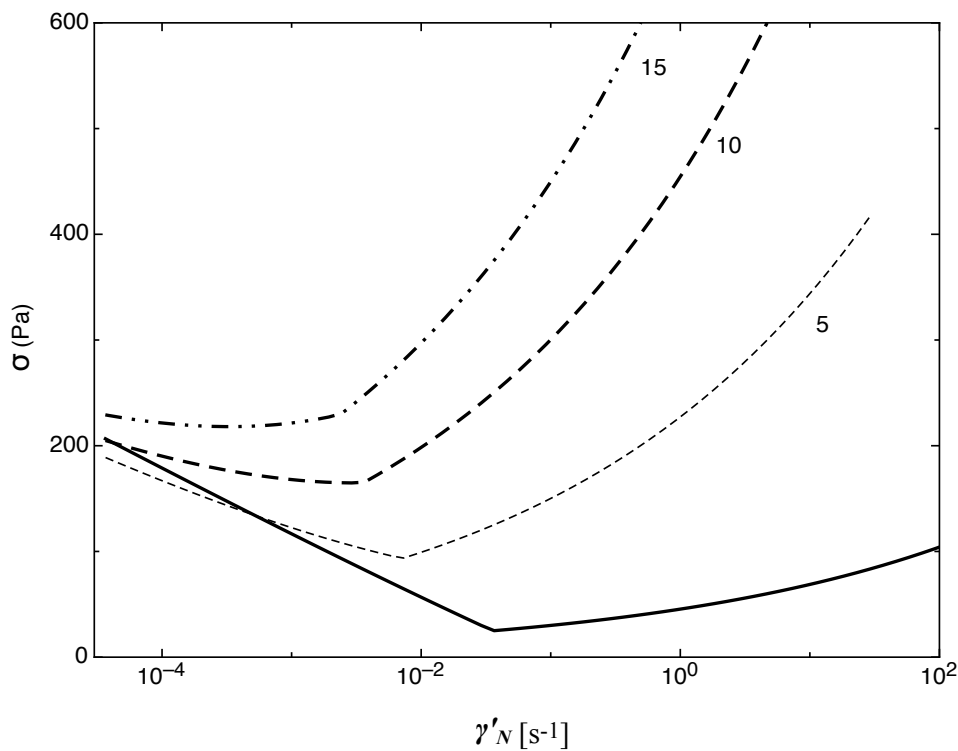

**Fig. 16.** Shows the effect of increasing the liquid phase viscosity has on the fitted flow curve from fig.1. The viscous term has been increased by the factor shown and the solid-phase stress has been shifted down the rate axis to keep its dependence on Pe unchanged.

*5.2 Adhesive yield at bounding surfaces versus cohesive yield.*

Colloidal particles can be caused to aggregate (coagulate or flocculation) by means of a number of different mechanisms, about ten in all. Most of them are indiscriminate inasmuch that they are likely to cause particles to adhere to bounding surfaces too. Because of this, cohesive suspension possess two yield stresses, and true yield stress associated with cohesive yielding within the suspension and an adhesive or wall yield stress, associated with slip at smooth boundaries, and dependent upon the nature of the surface and the suspension. That premature yield and slip at smooth boundaries are seen at all [2, loc. cit.] suggests that the adhesive stress tends to be smaller that the true yield stress and where the ratio of the two, adhesive to cohesive, has been determined, ratios of between 0.2 and 0.7 have been found. One might ask why? Not least because a consideration of interparticle forces in the Derjaguin approximation [16], suggests that the force between a particle and a wall should be twice the interparticle force, everything else being equal, which implies that the structure must be different at a wall, not unreasonably, perhaps. One possible scenario could even be that there is no "caging" at a wall and, if that was the case, then adhesive yield would always occurs at the bond strain. That being so then the adhesive to cohesive stress ratio would be given by $2\left(\dfrac{\gamma_{bond}}{\gamma_{cage}}\right)^{[1+m(Pe)]}$, according to eqn 3.8, which for $CaCO_3$ suspensions would give ca. 0.16 at low Pe, everything else being equal. The same calculation would however give somewhat larger values for truly colloidal particle, e.g. ca. 0.3 at 500 nm, say, and larger values still for nanoparticles. The possibility that the "cage" might be either non-existent or weaker at the wall motivates the performance of similar stress growth measurements using narrow gaps so as to promote premature yield at the outer cylinder. Such work is underway.

**5.3 - practical and processing implications.**

Just as controlled or partially controlled rheometric flows can be either rate or stress controlled, so can processing operations or flows be either kinematically controlled or pressure controlled, or mixed. In the case of kinematically controlled flows it is unlikely that the descending branch of the flow curve in fig. 1 would be ever be accessed for a material with the particle size of the $CaCO_3$, as the shear-rates are so very low, hence it would probably just behave as a simple Herschel-Bulkley fluid all practical applications of a kinematically controlled type. This can be confirmed by considering, for example, laminar pipe-flow in a circular cross-section pipe. It is then a straightforward calculation to show that even at a rather slow rate of say, ~ 10 s$^{-1}$ wall shear rate, adding the solid-phase stress $\sigma_s$ to the power-law viscous term V1 (cf. fig.1) would increase the pressure-drop needed to maintain the same volumetric flow rate as for V1 alone by just a few parts per million only. The same would not necessarily be true at much smaller particle size though perhaps, because of the inverse cubic dependence of Pe on particle size.

It is more difficult to say much about pressure-driven flows in the absence of a criterion for shear-banding, except that in complex flows a liquid having a flow curve of the type shown in fig. 1 would be expected to exploit any opportunity geometrical asymmetry might give it to bifurcate or bypass.

Many processes to which particulate suspensions are subjected involve flow together with changes in concentration. The latter can be unwanted, as in the laminar flow settling of cohesive suspensions [17], or the silting of estuaries, or they can be wanted, as in many solid-liquid separation operations. In most cases there is a combination of pressure-driven consolidation differential compression of the solid phase in rheological terms) and shear, be it an imposed flow, or simply yield and slip at boundaries like the walls of settling tanks. In order to model such processes it is necessary to know the ratio of shear and compressive strengths at all relevant volume-fractions. Since cohesive suspensions, strongly flocculated ones at least, are "ratchet poroelastic" [18] and strain-hardening in effect, the ratio of shear to compressive strength decays from unity at the gel-point to a lower asymptotic value at high volume-fraction. The latter is expected to proportional to the apparent yield strain in shear, defined as, $\gamma_Y^{app} = (\sigma_Y)/G$ [19, 20], and where the yield stress $\sigma_Y$ could either be the true yield stress or the adhesive, depending upon the context (e.g. whether any bounding surfaces are rough or smooth). It is of interest to ask how this apparent yield strain, which turns out to be convenient and compact way of the parameterizing the importance or otherwise of wall or shear effects in solid-liquid separation [19, 20], compares with the true yield strain. It should be fairly obvious that $\gamma_Y^{app} \geq \gamma_{bond}$ in the present terms, where the latter is the critical strain above which softening occurs: equal in the absence of the kind of strain hardening shown in fig. 10 and somewhat larger but of the same order of magnitude, with it. Hence the apparent yield strain might or might not be close to the actual yield strain, depending upon whether the latter is measured in controlled stress or controlled rate mode, and if the latter, upon Pe. The reason for discussing this aspect here is that the implication is that where either yield stress data or critical strain data (one being a proxy for the other) are needed for processing modeling of solid-liquid separation operations, it will be important to measure them in an appropriate way, supposing that rate-dependent strain softening and yield stress variation of the type shown in table 2 are anything like common as we suspect (a view that the widespread observation that the yield strain in creep is ~1 supports).

**6. Conclusions.**

It was shown in [1] that concentrated suspensions of coagulated 4.5 μm dia. CaCO3 particles show highly non-monotonic flow curves. The flow is stable in controlled rate testing, albeit intrinsically noisy. At any one shear-rate the material behaves as a Herschel-Bulkley liquid, but one with a shear-rate dependent yield stress – the solid phase stress, $\sigma_s$ here.

It was possible by means of plausible curve fits to decompose the stress into two additive parts, the solid phase stress, $\sigma_s$, plus a power-law viscous stress. It then was then possible to convert angular velocity (or equivalently, apparent Newtonian shear-rate at the inner tool) to true shear rate, whereas that cannot be done without such decomposition in the region where the stress decreases with rate.

Stress growth curves in step shear rate looked similar to those found by Petekidis et al. [4,5] qualitatively speaking. The separation between the softening strain and the strain of peak stress was however ca. two orders of magnitude larger as result of the larger particle size.

A plausible decomposition suggested that the viscous stress goes through a maximum at the second characteristic strain, whereas the solid phase stress $\sigma_s$ plateaued there. Both depend upon shear-rate above the softening strain, with $\sigma_s$ decreasing and the viscous stress increasing with increasing shear-rate. The net effect was that the total peak stress first decreased and increased with Pe, whereas Petekidis et al. [4,5] only saw an increase with their PMMA dispersions.

The Pe dependence of the solid phase stress could be characterised in terms of a Pe-dependent strain-softening exponent $m(Pe_0)$. At low $Pe_0$, where $m(Pe_0) > -1$, the stress $\sigma_s$ became fully developed at strain ~ 1, whereas at $Pe_0$ ~1 and above it either plateaued or peaked at the "bond" strain, even though the *total* stress still peaked at ~1 because of viscous stress growth. The net effect was that the total peak stress first decreases and increases with Pe, whereas Petekidis et al. [4,5] only saw an increase with Pe. One possible answer why they saw an increase only could be that solid phase and viscous stresses scaled very differently with particle size etc., causing the viscous stress to be dominant in their system. Their liquid phase viscosity was much higher too though and it could simply be that that alone was enough to obscure any rate dependence of their relatively small $\sigma_s$, as is suggested by fig. 16.

Overall then, do cohesive suspensions yield at the bond strain or the "cage" strain? The answer seems to be that they can do either in controlled rate testing or flow, depending upon Pe and, possibly, concentration. Controlled stress testing is another matter, although, there, softening at the bond strain followed by yield at strain ~ 1 has been seen widely. Furthermore, the same low Pe softening exponent of ca. 0.7 is obtained from creep. This pattern then is fully consistent with the idea of rate-dependent softening.

Just as controlled or partially controlled rheometric flows can be either rate or stress controlled, so can processing operations or flows be either kinematically controlled or pressure controlled (or, mixed). In the case of kinematically controlled flows it is unlikely that the descending branch of the flow curve in fig. 1 would be ever be accessed for a material with the particle size of the $CaCO_3$, as the shear-rates are very low, hence it would probably just behave as a simple Herschel-Bulkley fluid in most practical applications of a rate controlled type. The same would not necessarily be true at significantly smaller particle size.

Suggestions for further work, a systematic scan of particle size apart, include an attempt to decompose the total stress and work in the strain-softening region into elastic and dissipative parts. This could be done in principle by means of stress relaxation and by creep coupled with creep recovery. Also by LAOS perhaps, although a better use for LAOS, arguably, could be as a fingerprint of rate-dependent strain softening, since this should reverse the dependence of the phase-angle on frequency over a certain relevant range. The reason that it is important to disentangle the strain <1 region into elastic and dissipative parts in order to discover the true yield criterion underlying yield in simple shear, since this is as yet unknown, even if stored strain energy is a strong candidate. A fundamental criterion is needed in order to generalize to arbitrary deformations and loadings and to build constitutive equations.

There is a real need for simulations to probe yielding further. The discussion of microscopic or mesoscopic theories and particle level simulations has been avoided here deliberately. Not because there is not much interesting and excellent work described in the literature, far from it, but purely so as not to obscure or dilute the experimental picture that can be distilled from the data for $CaCO_3$, the aim here being to present an empirical model or scheme as a challenge to subsequent work. The reader might however wish to refer to the recent special issue of Journal of Rheology on colloidal gels [21] where can be found some very recent theoretical and simulation works.

**Appendix 1 – summary of key points regarding materials & methods from [1].**

A1.1 materials. The second material was commercial calcium carbonate (Omyacarb® 2-LU, Omya California Inc.) suspended in 0.01M of potassium nitrate and coagulated at the natural pH of 8.2 +/- 0.5. The weight-average mean particle size measured using a Malvern Mastersizer 2000 was 4.5 μm. The volume-fraction was 0.4, based on the manufacturers figure for the density of the particles of 2700 kg m$^{-3}$. The differential high-shear relative viscosity was of order 30, which again implies an effective volume-fraction much higher than the actual of 0.4.

A1.2 Methods Three different rheometers were employed, a TA Instruments AR-G2 rheometer in controlled stress and in controlled-rate mode a Rheometric Scientific$^{TM}$ ARES rheometer and a Haake VT550 instrument. It was not possible to use the same vane and cup in each instrument because the couplings were incompatible. The vane and cup dimensions used are given in Table 1:

**Table A1:** Dimensions of test geometries for AR-G2, ARES and Haake testing of the $CaCO_3$ suspension.

|  | AR-G2 | ARES | Haake |
|---|---|---|---|
| **Vane diameter (mm)** | 28 | 8 | 25 |
| **Cup diameter (mm)** | 142 | 34 | 75 |
| **Cup to vane diameter ratio** | 5.07:1 | 4.25:1 | 3:1 |
| **Vane length (mm)** | 42 | 16 | 50 |

**Acknowledgements.** Hui-En Teo was funded by a Brown Coal Innovation Australia Postgraduate Scholarship. Infrastructure support at Melbourne was provided by the Particulate Fluids Processing Centre, a Special Research Centre of the Australian Research Council.

**References.**

[1] Buscall, R., Kusuma, T.E., Stickland, A.D., Rubasingha, S., Scales, P.J., Teo, H-E, Worrall, G.L, The non-monotonic shear-thinning flow of two strongly cohesive concentrated suspensions, J. non-Newtonian Fluid Mech., DOI:10.1016/j.jnnfm.2014.09.010. (2014).

[2] Stickland, A.D., Kumar, A., Kusuma, T., Scales, P.J., Tindley, A., Biggs, S., & Buscall, R., The Effect of Vane-in-Cup Gap Width on Creep Testing of Strongly-Flocculated Suspensions, (revised submission to Rheol. Acta, January 2015). http://arxiv:1408.0069.

[3] Herschel, W.H.; Bulkley, R. (1926), Konsistenzmessungen von Gummi-Benzollösungen, Kolloid Zeitschrift **39**: 291–300.

[4] Koumakis, N. and G. Petekidis, Two step yielding in attractive colloids: transition from gels to attractive glasses, Soft Matter, **7**, 2456 (2011).

[5] Pham, K.N., G. Petekidis, D. Vlassopoulos, S. U. Egelhaaf, W. C. K. Poon and P. N. Pusey, Yielding behavior of repulsion- and attraction-dominated colloidal glasses, J. Rheology, **52**, 649 (2008).

[6] Yin G., and M. J. Solomon, Soft glassy rheology model applied to stress relaxation of a thermo-reversible colloidal gel, J. Rheol., **52**, 785–800 (2008).

[7] Chatzimina, M., G. Georgiou and A. Alexandrou, Wall Shear Rates in Circular Couette Flow of a Herschel-Bulkley Fluid, Appl. Rheol. **19** 32488:1-8 (2009).

[8] Onaka, S., Materials Transactions, **53**,8, 1547 -1548 (2012); Philos. Mag. Lett. **90** 633-639 (2010); loc. cit.

[9] Rehbinder P., Coagulation and thixotropic structures. Disc. Faraday Soc., **18**: 151-160. (1954).

[10] Uhlherr, P.H.T., J. Guo, C. Tiua, X.-M. Zhang, J.Z.-Q. Zhou, T.-N. Fang, The shear-induced solid–liquid transition in yield stress materials with chemically different structures, J. Non-Newtonian Fluid Mech. **125** 101–119 (2005).

[11] Gopalakrishnan,V. and C. F. Zukoski, Yielding behavior of thermo-reversible colloidal gels, Langmuir, **23** 8187–8193 (2007).

[12] Kumar A, Stickland AD, Scales PJ, Viscoelasticity of coagulated alumina suspensions, Korea-Australia Rheology J., **24:** 105-111 (2012).

[13] Bergstrom, L., Hamaker constants of inorganic materials, Adv. Colloid Interface Sci., **70** 125–169 (1996).

[14] Zwanzig R. and Mountain R. D. High-Frequency Elastic Moduli of Simple Fluids, J. Chem. Phys. **43** 4464-4471 **(1965).**

[15] Pasol, L. and X. Chateau, Elastic modulus of a colloidal suspension of rigid spheres at rest, C. R. Mecanique **336** 512-517 (2008).

[16] Derjaguin, B.V. Untersuchungen über die Reibung und Adhäsion, IV. Theorie des Anhaftens kleiner Teilchen, Kolloid Z. 69 (2): 155–164 (1934).

[17] http://www.patersoncooke.com/services/slurry/SLURRY_flow.pdf

[18] Kim, C., Y. Lui, A. Kuhnle, S. Hess, S. Viereck, T. Danner, L. Mahadevan

and D.A. Weitz., Gravitational Stability of Suspensions of Attractive Colloidal Particles, Phys. Rev. Lett., 2007, 99, 028303.

[19] Lester, D.R., Buscall, R., A.D. Stickland, P. J. Scales, Wall Adhesion and Constitutive Modelling of Strong Colloidal Gels, Rheol., 58(5):1247–1276, (2014).

[20] Lester, D.R. & R. Buscall, Correction of Wall Adhesion effects in Batch Settling of Strong Colloidal Gels, J. non-Newtonian Fluid Mech., (submitted August 2014). http://arxiv.org/abs/1409.8397.

[21] 'Special Issue on Colloidal Gels', J. Rheol. **58**(5) 1085-1618 (2014).

[22] Dispersion Polymerisation in Organic Media, Barrett, K.E. (ed.), Wiley (1974).

[23] 'Colloidal Dispersions', Russel, W.B., Saville, D.A. and Schowalter, W.R., Cambridge University Press, Cambridge (1989), ISBN 0 521 34188 4, fig. 8.4, page 271.